\documentclass[twocolumn]{aastex61}

\pdfoutput=1 
\usepackage{amsmath,amstext}
\usepackage[T1]{fontenc}
\usepackage{apjfonts} 
\usepackage[figure,figure*]{hypcap}
\usepackage{xspace}
\usepackage{gensymb}

\newcommand{\Ha}{H$\alpha$\xspace}
\newcommand{\pp}{P96949\xspace}

\defcitealias{Poggianti2017a}{Paper I}
\defcitealias{Fritz2017}{Paper III}
\defcitealias{Gullieuszik2017}{Paper IV}

\shorttitle{Tidal Dwarf Galaxy in a merging system}
\shortauthors{B. Vulcani et al.}

\begin{document}

\title{
GASP VIII. Capturing the birth of a Tidal Dwarf Galaxy in a  merging system at \lowercase{z}$\sim$0.05}

\author{Benedetta Vulcani}
\affiliation{School of Physics, University of Melbourne, VIC 3010, Australia}
\affiliation{INAF- Osservatorio astronomico di Padova, Vicolo Osservatorio 5, IT-35122 Padova, Italy}
\author{Alessia Moretti}
\affiliation{INAF- Osservatorio astronomico di Padova, Vicolo Osservatorio 5, IT-35122 Padova, Italy}
\author{Bianca M. Poggianti}
\affiliation{INAF- Osservatorio astronomico di Padova, Vicolo Osservatorio 5, IT-35122 Padova, Italy}
\author{Giovanni Fasano}
\affiliation{INAF- Osservatorio astronomico di Padova, Vicolo Osservatorio 5, IT-35122 Padova, Italy}
\author{Jacopo Fritz}
\affiliation{Instituto de Radioastronom\'ia y Astrof\'isica,
UNAM, Campus Morelia, A.P. 3-72, C.P. 58089, Mexico}
\author{Marco  Gullieuszik}
\affiliation{INAF- Osservatorio astronomico di Padova, Vicolo Osservatorio 5, IT-35122 Padova, Italy}
\author{Pierre-Alain Duc}
\affiliation{Universit\'e de Strasbourg, CNRS, Observatoire astronomique de Strasbourg, UMR 7550, F-67000 Strasbourg, France}
\author{Yara Jaff\'e}
\affiliation{European Southern Observatory, Alonso de Cordova 3107, Vitacura, Casilla 19001, Santiago de Chile, Chile}
\author{Daniela Bettoni}
\affiliation{INAF- Osservatorio astronomico di Padova, Vicolo Osservatorio 5, IT-35122 Padova, Italy}

\begin{abstract}
Within the GAs Stripping Phenomena in galaxies with MUSE (GASP) sample, we identified an ongoing 1:1 merger between two galaxies and the consequent formation of  a tidal dwarf galaxy (TDG). The  system is observed at $z = 0.05043$ and is part of a poor group. 
Exploiting the exquisite quality of the MUSE/VLT data, we  present the spatially resolved kinematics and physical properties of gas and stars of this object and describe its evolutionary history.
An old (luminosity weighted age $\rm \sim 2\times 10^9 \, yr$), gas poor,  early-type-like galaxy  is merging  with a  younger (luminosity weighted age $\rm \sim 2.5\times 10^8 \, yr$), gas rich,  late-type galaxy . The system has a quite strong metallicity gradient, indicative of an early-stage phase. Comparing the  spatial extension of the star formation at different epochs, we  date the beginning of the merger between $\rm 2\times 10^7  yr <t<5.7\times 10^8  yr$ ago.  The gas kinematic pattern reflects that of the late-type object and is distorted in correspondence to the location of the impact \deleted{, while the northern regions had not time  to be noticeably influenced yet}. The stellar kinematic instead is  more chaotic, as expected for mergers. 
The gas redistribution in the system  induces high levels of star formation between the two components,  where we indeed detect the birth of the TDG. 
This stellar structure has a mass of $\sim 6\times 10^9 M_\odot$, a radius of $\rm \sim 2 \, kpc$ and, even though it has already accreted large quantities of gas and stars, it is still located within the disk of the progenitor, is characterized by a high velocity dispersion, indicating that it is still forming, is dusty and has  high levels of star formation (SFR$\sim 0.3 M_\odot \, yr^{-1}$). 
This TDG is originated in an early-stage merger, while these  structures usually form in more evolved systems. 
\end{abstract}


\section{Introduction}\label{sec:intro}
According to the hierarchical structure formation paradigm \citep{Blumenthal1984,  Freedman2001, Efstathiou2002, Pryke2002, Spergel2007}, smaller systems merge to form progressively larger ones \citep{White1978,Searle1978}. Mergers therefore are fundamental features in galaxy evolution.

Galaxy merging is expected to have a number of consequences on galaxy evolution. Its exact effect depends on a wide variety of parameters such as collision angles, speeds, relative size/composition and environment. Typically, it drives strong star formation episodes \citep{Barnes1991,Mihos1996,  Larson1978, Kennicutt1998a, Elmegreen2011}, contributes to and regulate the growth of black holes \citep{Kauffmann2003, DiMatteo2005, Cox2006, Schawinski2009, Combes2003}, and produces morphological transformations \citep{Toomre1977, Mihos1996}.

During a major merger (with mass ratios from 1:1 to about 3:1), typical star formation rates (SFRs) are less than $\rm 10^2 \, M_\odot \, yr^{-1}$  \citep{Moster2011, Hirschmann2012}, but  can reach peaks of $\rm 10^3 M_\odot \, yr^{-1}$, depending on the gas content of each galaxy and its redshift \citep[e.g.,][]{Brinchmann2004, Ostriker2011}. Recent high-resolution models forecast that extended star formation is important in the early stages of the merger, while  nuclear starbursts will occur in advanced stages \citep{Teyssier2010,  Hopkins2013, Renaud2015}.
The fast exhaustion of gas ($<$ few Gyr) during the merger induces the quenching of the  star formation and the consequent redistribution of the angular momentum  and the violent relaxation in the stellar component, with the transformation of galaxies into an early-type system.

The higher star formation efficiency, along with tidal fields, gravity torques leading to central inflows and outer outflows, increased turbulence, and high Jeans masses  induce the formation of a large variety of stellar structures, such as kinematically decoupled cores, super star cluster (SSCs) and tidal dwarf galaxies (TDGs) \citep[e.g.,][]{Bournaud2010}. 

Simulations by \cite{Elmegreen1993} show that SSCs  form from local gravitational instabilities. They can have a similar origin in tidal tails and in isolated systems. 
These giant molecular clouds  form with very high pressures and high gas densities \citep{Li2004, Bournaud2008} and can reach a mass of $\sim 10^6 M_\odot$. The star formation efficiency is  expected to be very high, so that these SSCs 
 may likely remain bound after the expulsion of gas by the first generations of supernovae, and evolve into globular clusters \citep{Bournaud2008, Recchi2007, Renaud2015}.

TDGs form instead only  in  interacting galaxies, where some disk material become bound and  star-forming as a consequence of a ``pile-up'' in some particular regions. A dense region in a disk (proto-cloud) can be moved outwards as part of a tidal tail. There it can remain bound instead of fragmenting into an unbound complex of several smaller pieces, because of the increased velocity dispersion \citep{Elmegreen1993, Duc2004}. 
These  massive ($\sim 10^8 M_\odot$)  star forming structures are made-up mostly of gas and new stars formed locally. The pre-existing stars, present in the parent galaxies before the collision, generally have a too high random velocities, and escape these newly-formed system.
TDGs are therefore kinematically decoupled from their parent galaxy, even though gravitationally bound \citep{Bournaud2007, Duc2012}.

Many TDGs form around or above the critical mass for dwarf galaxies to survive the star formation feedback produced by their initial starburst \citep{Dekel1986, Boquien2007}. 
Simulations by \cite{Bournaud2006}  show that a significant fraction of TDGs can survive several Gyrs, orbiting as satellite galaxies around the merger remnant.  
They may thus contribute to the population of dwarf satellites 
\citep{Okazaki2000, Metz2007}.

According to the merger simulations by \cite{Bournaud2008, Chapon2013}, local gravitational instabilities  and  the pile-up of large amounts of gas can happen at the same time. A single merger rarely form more than a couple of  objects through the pile-up mechanism, while many more through instabilities. 

Integral field spectroscopy can provide relevant information to characterize merging systems and the resulting stellar structures, since it allows to spatially resolve the properties of the components and to characterize the extent of star formation, and how/when it is produced during the merger event. The observational characterization of star formation in mergers at different stages is also necessary to test the validity and put constraints on merger simulations. 

In the last few years, a number of specific interacting/merging systems in the local universe have been observed with such techniques at optical wavelengths \citep[e.g.,][]{Wild2014,  Fernandez2015, Cortijo2017a, Cortijo2017b, Cortijo2017c}. \cite{Cortijo2017b} have also characterized the star cluster properties in an early-stage merger. All these systems are at $z\leq 0.04$ and have been drawn from the CALIFA \citep{Sanchez2012} and MaNGA \citep{Bundy2015} surveys, which are characterized by a spatial sampling of 1$^{\prime\prime}$/pixel  and $2\farcs 5$/pixel and a resolution of $\rm \sim 80 \, km \, s^{-1}$ and  $\rm \sim 70 \, km \, s^{-1}$, respectively.    

A number of studies based on integral field spectroscopy have also given us some insight into the properties of dwarf galaxies \citep{Izotov2006, Lagos2009, Lagos2012, Lagos2014, James2009, James2010, James2013a,James2013b}. 

This paper  presents the analysis of a merging system at $z\sim 0.05$, observed with  the integral-field spectrograph MUSE mounted at the VLT, which is currently the most powerful instrument of this kind, having a spatial sampling of$\sim0 \farcs 2$/pixel and a resolution of $\rm \sim 50 \, km \, s^{-1}$. 
The exquisite data quality allows us to detect the formation of a TDG candidate in correspondence of the impact between the two merging components.  This is the highest redshift TDG detected up to date.  To our knowledge, only \cite{Fensch2016} have exploited the  unprecedented spatial resolution of MUSE observations to characterize the physical conditions where a dwarf galaxy formed in a giant collisional  {\sc Hi} ring at a distance of 63.1 Mpc.

The system has been observed in the context of GASP\footnote{\url{http://web.oapd.inaf.it/gasp/index.html}} (GAs Stripping Phenomena in galaxies with MUSE), an ongoing ESO Large programme granted 120 hours of observing time with  MUSE. The program is aimed at characterizing where, how and why gas can get removed from galaxies.
\citet[][Paper I]{Poggianti2017a} presents a complete description of the survey strategy, data reduction and analysis procedures. 

The galaxy we discuss here was selected for presenting a B-band morphological asymmetry. 
Only the integral field spectroscopy allowed us to unambiguously identify the  merging event. Note that this system is so far the only merger detected in the survey, indicating that usually signatures of mergers and stripping are distinct. 

Throughout all the papers of the GASP series, we adopt a \cite{Chabrier2003} initial mass function (IMF) in the mass range 0.1-100 M$_{\odot}$. The cosmological constants assumed are $\Omega_m=0.3$, $\Omega_{\Lambda}=0.7$ and H$_0=70$ km s$^{-1}$ Mpc$^{-1}$. This gives a scale of 0.985 kpc/$^{\prime\prime}$ at the redshift of the merging system, which is $z = 0.05043$.

\section{The target}\label{sec:data}

\begin{figure}
\centering
\includegraphics[scale=0.4]{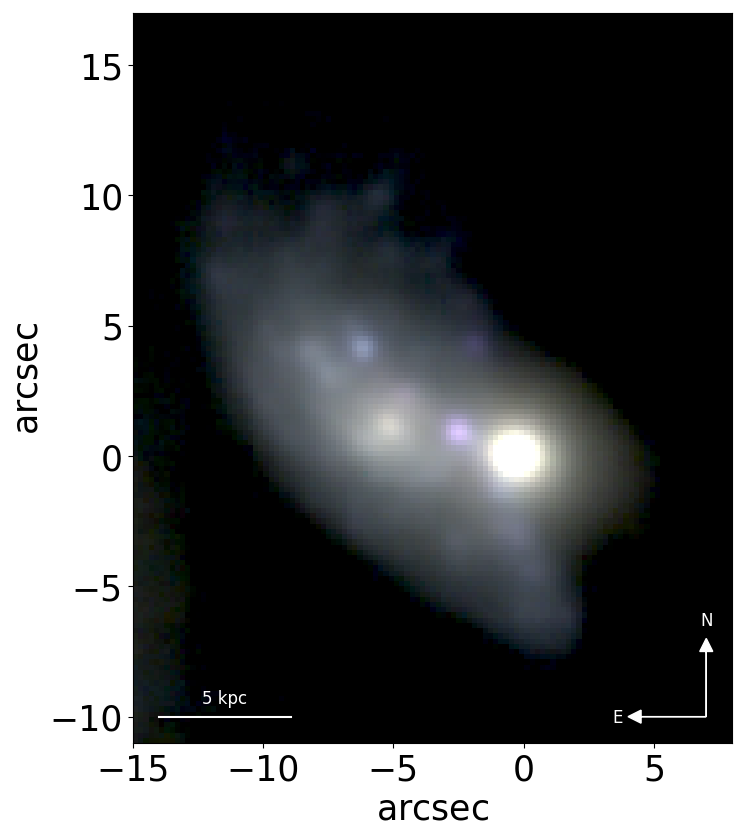}
\caption{RGB image of \pp. The reconstructed $g$, $r$, $i$ filters from the MUSE cube have been used. 
North is up, and east is left.  In this and all plots (0,0) corresponds to the peak of the  continuum underlying \Ha flux (see Fig. \ref{fig:white}). \label{fig:rgb_image} }
\end{figure}

\begin{figure}
\centering
\includegraphics[scale=0.3]{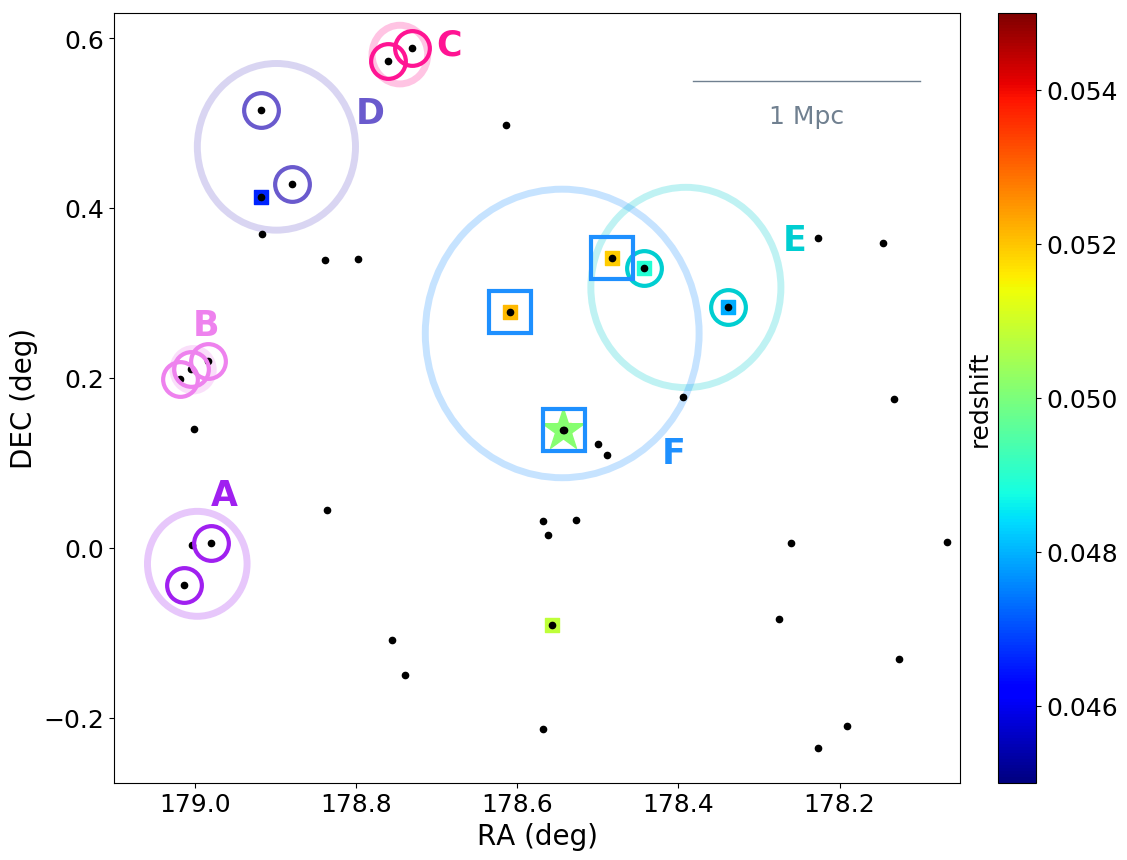}
\caption{Position on the sky of galaxies around \pp, indicated with a green  star,  in the redshift range  $0.01<z<0.1$ (black dots). Galaxies in the redshift range $0.045<z<0.055$ are color-coded according to their redshift (squares). Galaxies belonging to the group of \pp (group F) are highlighted with open squares; galaxies belonging to other groups are highlighted with open circles (see  \citealt{Tempel2012} and Tab.\ref{tab:groups} for details). Shaded circles indicate the virial radius of the groups. The scale in the upper  right corner shows 1 Mpc at the redshift of \pp. \label{fig:env} }
\end{figure}

The target, \pp (RA:11:54:10.27, DEC:+00:08:19.3),  is  drawn from the Millennium Galaxy Catalog \citep{Liske2003,Driver2005} and selected by \cite{Poggianti2016} from the Padova Millennium Galaxy and Group Catalogue \citep[PM2GC,][]{Calvi2011}.

Figure \ref{fig:rgb_image} shows a color composite image obtained combining the reconstructed $g$, $r$ and $i$ filters from the MUSE datacube. The irregular shape of the galaxy immediately stands out. Two distinct components are visible: a bright, circular one dominates on the West side of the galaxy, while a more diffuse, fainter one is  centered  $5\farcs 3$ North-East from the other one.  These two components  most likely correspond to two merging galaxies, but only the integral field spectroscopy can reveal whether they are physically associated.
A very bright knot in between the components already stands out for its peculiar colors.

\begin{table}
\caption{Properties of the groups around \pp \label{tab:groups}}
\centering
\setlength{\tabcolsep}{2pt}
\begin{tabular}{rrrrrrr}
\hline
  \multicolumn{1}{c}{IDcl} &
  \multicolumn{1}{c}{Ngal} &
  \multicolumn{1}{c}{z} &
  \multicolumn{1}{c}{RAJ2000} &
  \multicolumn{1}{c}{DEJ2000} &
  \multicolumn{1}{c}{Rvir} &
  \multicolumn{1}{c}{$\sigma$} \\
  \multicolumn{1}{c}{} &
  \multicolumn{1}{c}{} &
  \multicolumn{1}{c}{} &
  \multicolumn{1}{c}{(J2000)} &
  \multicolumn{1}{c}{(J2000)} &
  \multicolumn{1}{c}{(Mpc)} &
  \multicolumn{1}{c}{($\rm km \, s^{-1}$)} \\

\hline
  A & 2 & 0.07853 & 178.99687 & -0.01874 & 0.224 & 38.5\\
  B & 3 & 0.07944 & 179.00206 & 0.20978 & 0.091 & 248.3\\
  C & 2 & 0.07978 & 178.74551 & 0.58081 & 0.127 & 135.6\\
  D & 2 & 0.06184 & 178.89868 & 0.47222 & 0.285 & 9.8\\
  E & 2 & 0.04844 & 178.39067 & 0.30664 & 0.273 & 198.7\\
  F & 3 & 0.0519 & 178.54411 & 0.25248 & 0.420 & 60.0\\
\hline\end{tabular}
 \tablecomments{Data taken from \citet{Tempel2012}. }
\end{table}

\pp is part of a small group of three objects.  \cite{Tempel2012} computed a velocity dispersion $\sigma$=60 km s$^{-1}$. The other two group members are at 9.$^\prime$2 and 12.$^\prime$70 towards North. Figure \ref{fig:env} shows the spatial distribution of all galaxies around \pp with measured redshift in the range $0.01<z<0.1$. Redshifts are taken from the MGCz \citep{Driver2005} and SDSS-DR9 \citep{Ahn2012}. \cite{Tempel2012} identified  five other groups in this area, whose position is highlighted in Fig. \ref{fig:env} and whose properties are listed in Tab.\ref{tab:groups}. 
The relative positions, velocity dispersions and redshift differences  between all  these groups  and those of that of \pp (group ``F'') suggest that group-group interactions are hardly able to affect the group members.

\section{Data}
\subsection{Observations and data reduction}
Following the GASP strategy, \pp was observed in
service mode with the MUSE spectrograph, mounted at the Nasmyth focus of the UT4 VLT, at Cerro Paranal in Chile. It was observed on 21/02/2017, with photometric conditions; the seeing at 650nm (measured on telescope guide star)
remained below 0$\farcs$9 during the whole observing block.  A total of four 675 seconds exposures were taken with the Wide Field Mode. 

The data reduction process for all galaxies in the GASP survey is presented in  \citetalias{Poggianti2017a}.  

\subsection{Data analysis}\label{sec:analysis}
The procedures  used to analyze all galaxies of the GASP survey are extensively presented in \citetalias{Poggianti2017a}. 
 We corrected the reduced datacube for extinction due to our Galaxy. We assumed the extinction law from \cite{Cardelli1989} and used the extinction value computed at the galaxy position \citep{Schlafly2011}.

We analyzed the most prominent emission lines in the spectrum by exploiting the {\sc kubeviz}  \citep{Fossati2016} code, which yields total fluxes and the kinematic properties of the gas. As first step, we average filtered the datacube in the spatial direction with a 5$\times$5 pixel kernel, corresponding to 1$^{\prime\prime}\sim 0.99$ kpc at the galaxy redshift \citepalias[see][for details]{Poggianti2017a}. 
In the MUSE wavelength range the typical spectral dispersion of 1.25 \AA{} pixel$^{-1}$ translates to a velocity scale of 25 km s$^{-1}$ pixel$^{-1}$. The average FWHM resolution is 2.51 \AA{}, equivalently to 110 km s$^{-1}$ (or 53 km s$^{-1}$ pixel$^{-1}$). Details on the methods can be found in \citetalias{Poggianti2017a}.

We derived the stellar kinematic from the analysis of the characteristics of absorption lines, using the  Penalized Pixel-Fitting (pPXF) software \citep{Cappellari2012}, which works
in Voronoi binned regions of given S/N \citep[10 in this case; see][]{Cappellari2012_v}.
We fitted observed spectra with the stellar population templates by \cite{Vazdekis2010}.
We further smoothed the value of the stellar radial velocity  using the two-dimensional local regression techniques (LOESS) as implemented in the Python code developed by M. Cappellari.\footnote{\url{http://www-astro.physics.ox.ac.uk/~mxc/software}}Details on the methods can be found in \citetalias{Poggianti2017a}.

We obtained the spatially resolved properties of the stellar populations and the correction for underlying absorption to measure total emission line fluxes running the spectral fitting code  {\sc sinopsis} \citep[Paper III]{Fritz2017}.  
Briefly, it combines different simple stellar populations  spectra to reproduce the observed equivalent widths of the most prominent absorption and emission lines, and the continuum in various  bands. 
The code uses the latest SSP model from S. Charlot \& G. Bruzual (2017, in preparation).
{\sc sinopsis} produces a best-fit model datacube for the stellar-only component and maps of stellar mass, average star formation rate and total mass formed in four age bins,  luminosity-weighted and mass-weighted stellar ages. 
To calculate the total mass in a given spaxel, we sum the masses in the four main age bins. Due to the code characteristics, when the spectra have a low signal-to-noise, {\sc sinopsis} tends to include an unnecessary small percentage of old (t>$5.7\times 10^8$) stars.
To be conservative, we neglect the contribution of stars older than $5.7\times 10^8$ yr in low S/N spectra (S/N$<$3) when their contribution to the stellar continuum luminosity is less than 3\%. The entire  contribution of young stars, instead, is taken into account, given the fact that it is estimated from the emission lines, which are more reliable features.

We then corrected the emission-line, absorption-corrected fluxes  for extinction by dust internal to the galaxy. We obtained the map of the dust extinction  A$_V$  from the 
absorption-corrected
Balmer decrement in each spaxel. As described in \citetalias{Poggianti2017a}, we assumed an intrinsic \Ha/H$\beta$ ratio equal to 2.86 and adopt the \cite{Cardelli1989} extinction law. 
The A$_V$ map has been calculated only for spaxels where the S/N on the H$\alpha$ and H$\beta$ lines is larger than 3 and the ratio of the two lines is larger than the assumed 2.86 value for the Balmer decrement. 

We used the line fluxes to produce line-ratio diagnostic diagrams \citep[BPT,][]{Baldwin1981} that allow to investigate the origin of the gas ionization and distinguish between regions photoionized by hot stars and regions ionized by shocks, LINERs and AGN. Only spaxels with a $S/N> 3$ in all the emission lines involved are taken into account. 

We calculated the gas metallicity for
each star forming spaxel using the pyqz Python
module7 \citep{Dopita2013} v0.8.2; we obtained the 
$12 + \log(O/H)$ values by interpolating
from a finite set of diagnostic line ratio grids computed
with the MAPPINGS code 
(see \citetalias{Poggianti2017a}).
As discussed in  detail by \cite{Kewley2008}, the systematic errors introduced by modeling inaccuracies are of the order of $\sim$0.1-0.15 dex. Instead, discrepancies among the various calibrations based on photoionization models can reach values of up to 0.2 dex. 

We computed the SFR of each spatial element from the \Ha luminosity corrected for dust and stellar absorption, following the \cite{Kennicutt1998a}'s relation. 
We finally derived the total SFR as the sum of the dust- corrected \Ha fluxes in each spaxel with a S/N(\Ha)$> 3$.

\section{Results}\label{sec:results}
\begin{figure}
\centering
\includegraphics[scale=0.36]{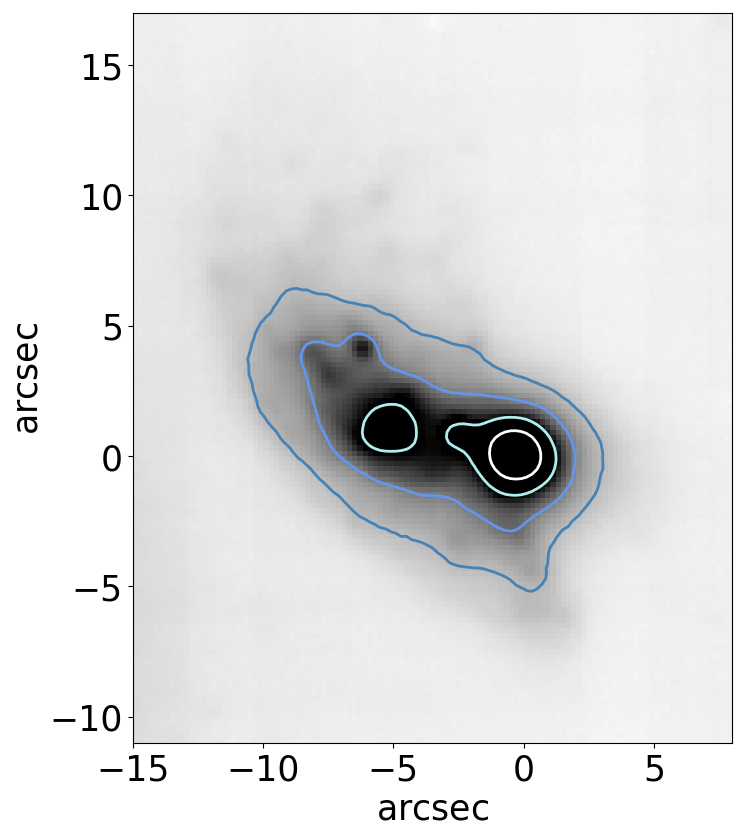}
\caption{MUSE white-light image of \pp.  Contours are logarithmically spaced isophotes of the continuum underlying \Ha down to a surface brightness $\rm 2.5\times 10^{-18} \, erg \, s^{-1} \, cm^{-2} \,$\AA{}$\rm ^{-1} \, arcsec^{-2}$.  \label{fig:white} }
\end{figure}

Figure \ref{fig:white} shows the white-light image from MUSE, i.e. the image obtained integrating the light across the entire wavelength range (4750-9350 \AA{}). The irregular shape of the galaxy is well visible, along with  traces of debris to the North  and the few bright knots distributed across the galaxy. The stellar isophotes reveal the presence of two distinct peaks in the light distribution, indicating the presence of two components in the system. 

\subsection{Spatially resolved gas properties}
\begin{figure}
\centering
\includegraphics[scale=0.35]{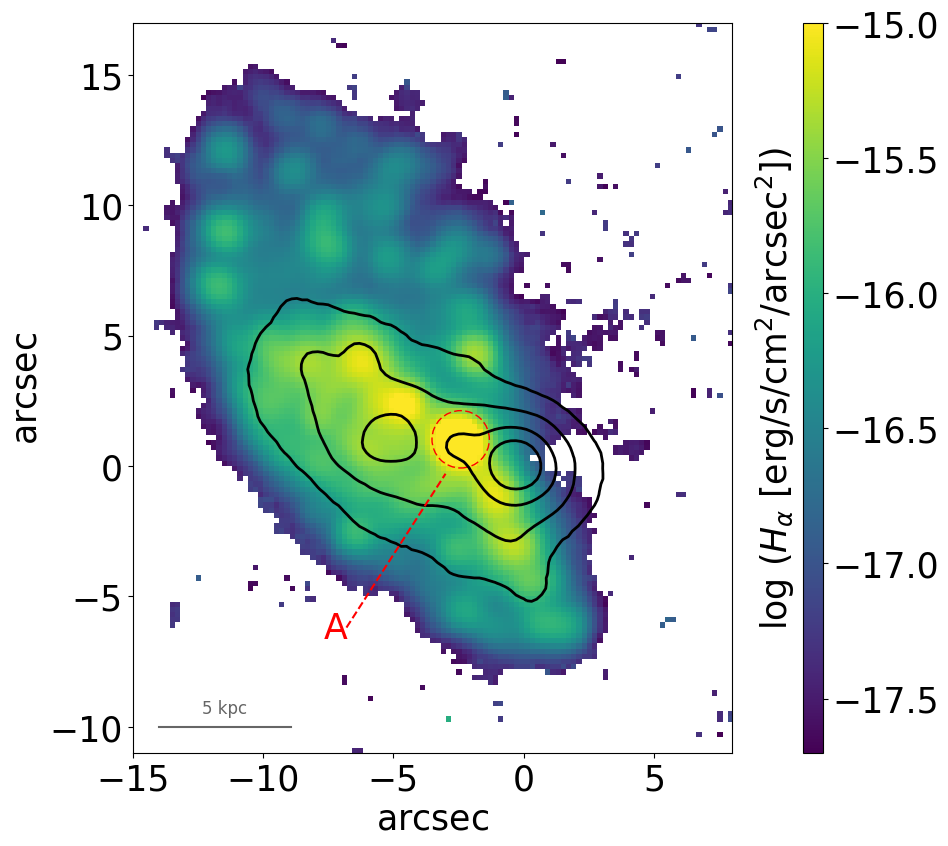}
\caption{MUSE \Ha luminosity map for spaxels with \Ha S/N$> 4$, corrected for Galactic extinction, but uncorrected for stellar absorption and intrinsic dust extinction.  Contours are continuum isophotes as in Fig.\ref{fig:white}. The position of the brightest region is highlighted in red and labeled as ``A''.\label{fig:Ha}}
\end{figure}

The image of the galaxy in the \Ha light shown in Fig. \ref{fig:Ha} unveils that the gas is much more extended towards North than the stellar continuum, and less in the Western part of the galaxy. The peaks in the stellar continuum do not coincide with peaks in the \Ha light. 
The \Ha map is characterized by many bright and round \Ha knots. These have  different sizes depending on their spatial location: those outside of the stellar contours are smaller, the few in the center of the galaxy are larger. A particularly bright and large knot stands out at the position ($x=-2.5^{\prime\prime}$, $y= 1^{\prime\prime}$). Given its peculiarity, in what follows we will pay particular attention to the properties of gas and stars in that region, which we refer as ``knot A''.

\begin{figure}
\centering
\includegraphics[scale=0.36]{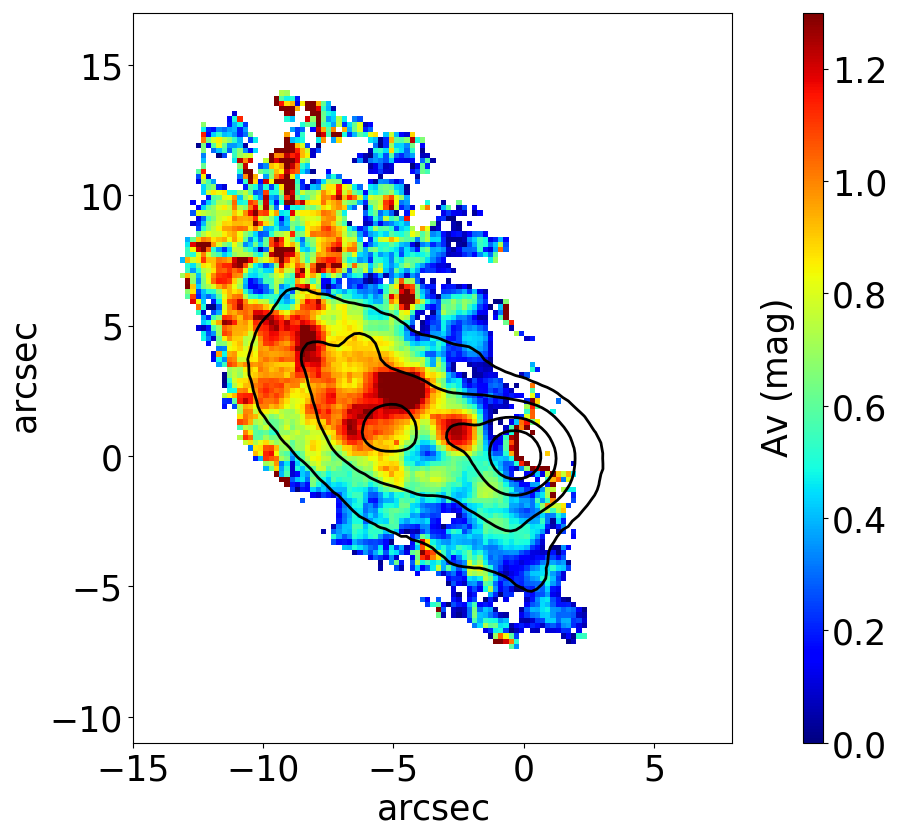}
\caption{$A_V$ map. Only spaxels with a S/N(\Ha)> 3  are shown.  Contours are continuum isophotes as in Fig.\ref{fig:white}. \label{fig:av}}
\end{figure}

As mentioned in Sec.\ref{sec:analysis}, the $\rm H\alpha$ fluxes  presented in Fig. \ref{fig:Ha} are corrected for extinction by dust internal to the galaxy.
The $A_V$ map (Fig. \ref{fig:av}) shows that the dust is distributed across \pp in a non-homogeneous way: in the South-West region the extinction  is very low ($A_V$<0.5 mag), while the Central-East and Northern sides are  dustier ($A_V\sim$1.2 mag).  Very dusty regions ($A_V\sim$1.4 mag) are located at the positions ($x=-5^{\prime\prime}$, $y= 3^{\prime\prime}$) and in correspondence of knot A.

The  map of \Ha, along with those of  H$\beta$, [OIII] 5007 \AA{}, [OI] 6300 \AA{}, \Ha, [NII] 6583 \AA{}, and [SII] 6716+6731 \AA{}, is used to determine the  main ionizing source at each position.
All the diagnostic diagrams used are concordant in finding that  young stars produce the ionized gas and in excluding the presence of AGN in the galaxy center. 
This is in agreement with previous classifications found in the literature for the same galaxy \citep[e.g.,][]{Veron2010}. 
\begin{figure}
\centering
\includegraphics[scale=0.36]{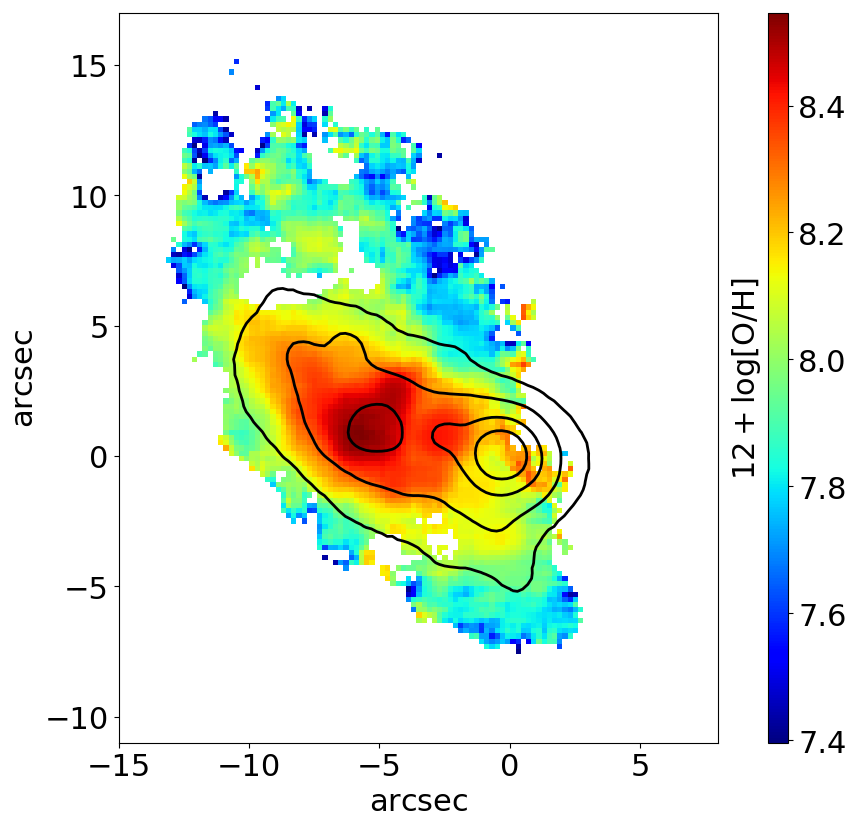}
\caption{Metallicity map from the ionized gas. Contours are continuum isophotes as in Fig.\ref{fig:white}. \label{fig:metallcity}}
\end{figure}

The gas metallicity of the ionized gas is shown in Fig. \ref{fig:metallcity}. The distribution is characterized by a quite strong gradient, varying over almost 1 dex in $12+\log[O/H]$, with the highest metallicity regions located at the position of the nucleus of the easternmost component. This does not correspond to a peak in neither \Ha nor $A_V$ (see Figures \ref{fig:Ha} ans \ref{fig:av}, respectively), but it does correspond to the center of the fainter galaxy, as traced by the stellar continuum. A secondary peak is observed very close to the primary, in the position $x=-5^{\prime\prime}$, $y= 3^{\prime\prime}$, where instead a star-forming region of particularly high \Ha brightness  and with rather high dust extinction is identified. Knot A is also characterized by relatively high values of metallicity.

\subsection{Gas and stellar kinematics}
\begin{figure*}
\centering
\includegraphics[scale=0.38]{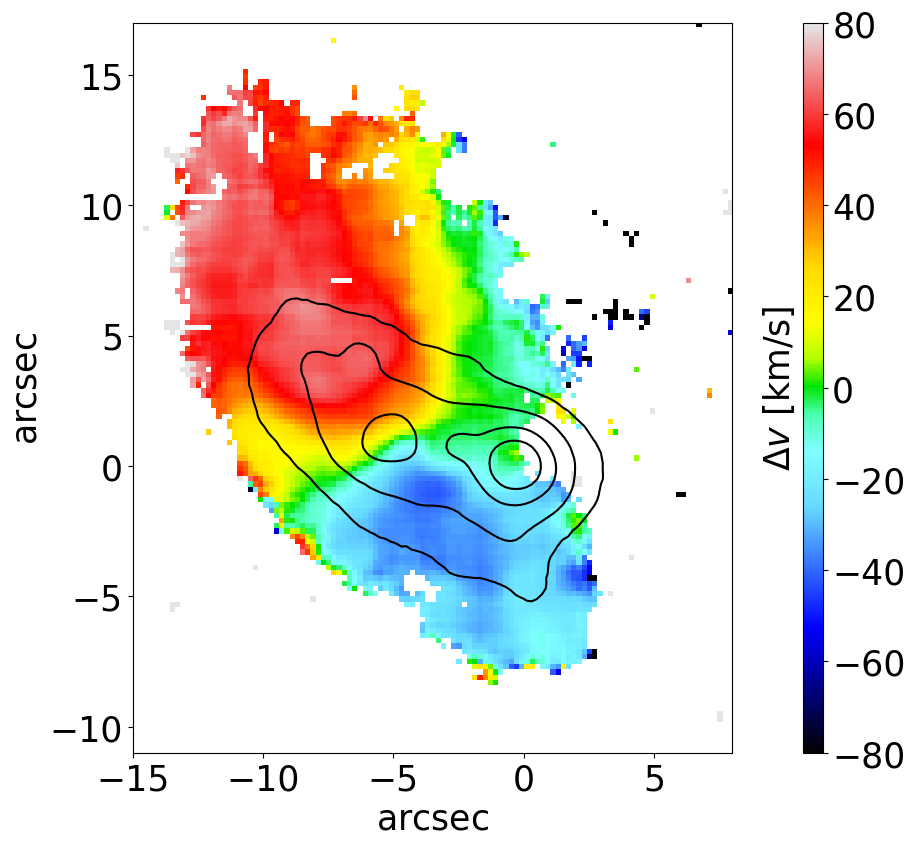}
\includegraphics[scale=0.38]{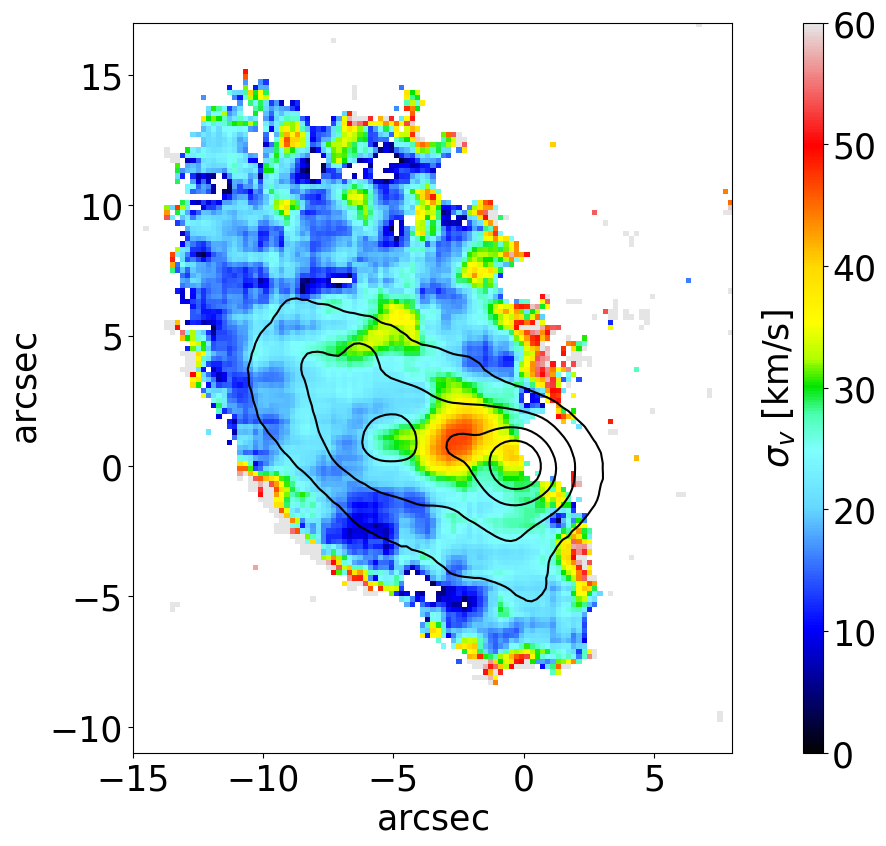}
\caption{Gas kinematics maps for 5$\times$5 spaxels with S/N \Ha > 4. Left: \Ha velocity. Right:  \Ha velocity dispersion.  Contours are continuum isophotes as in Fig.\ref{fig:white}. \label{fig:vel_gas}}
\end{figure*}

Figure \ref{fig:vel_gas} shows the velocity field of the gas component. Assuming as zero point of the velocity the redshift of the galaxy, the gas rotates along the South-East/North-West direction and spans the velocity range $-50<v (\rm km \, s^{-1}) <80$. The median error on the velocities is 1.4 km s$^{-1}$. 
The gas presents a regular rotation in the outskirts, with positive velocity values in the Northern region, negative velocity values in the Southern one. In the center, the locus of the zero velocity is highly distorted and forms a ``{\it y}'' shape. The gas velocity dispersion is overall quite low ($\sigma_{gas}=10-20 \, \rm km \, s^{-1}$, with a median error of 3.5 km s$^{-1}$), 
indicative of a dynamically cold medium, except 
at the position of the knot A, 
where it reaches peaks of $\sim 50 \, \rm km \, s^{-1}$. Similar values are found  at the edges of \pp, but hey are probably due to the low S/N ratio of these regions. 

\begin{figure*}
\centering
\includegraphics[scale=0.38]{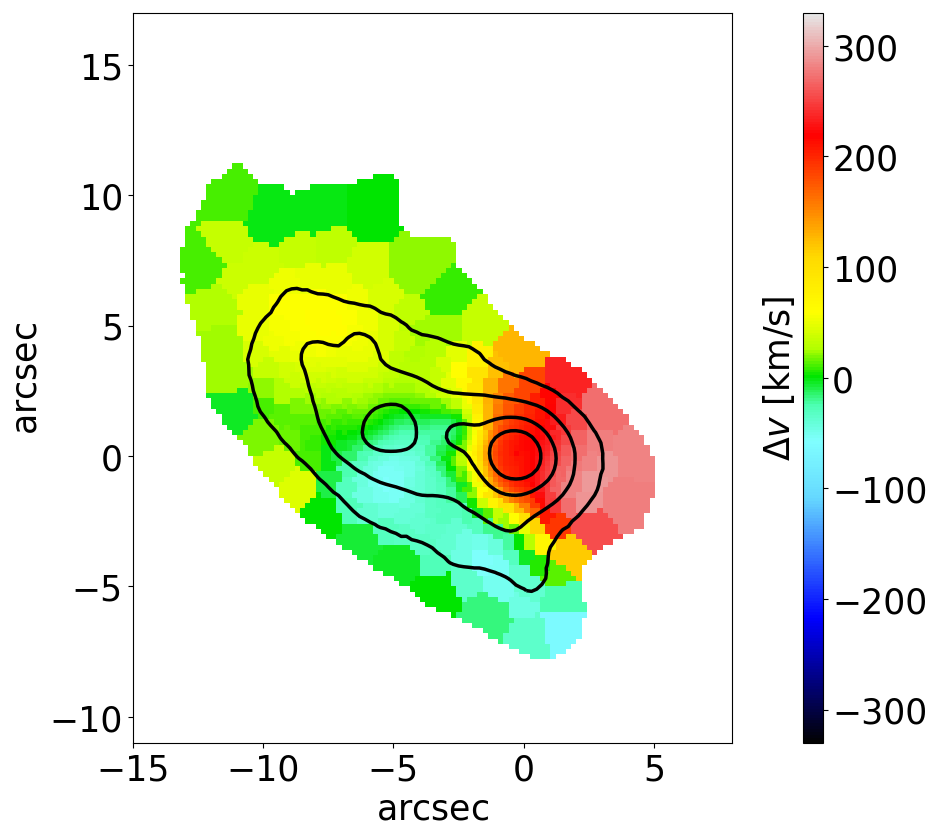}
\includegraphics[scale=0.38]{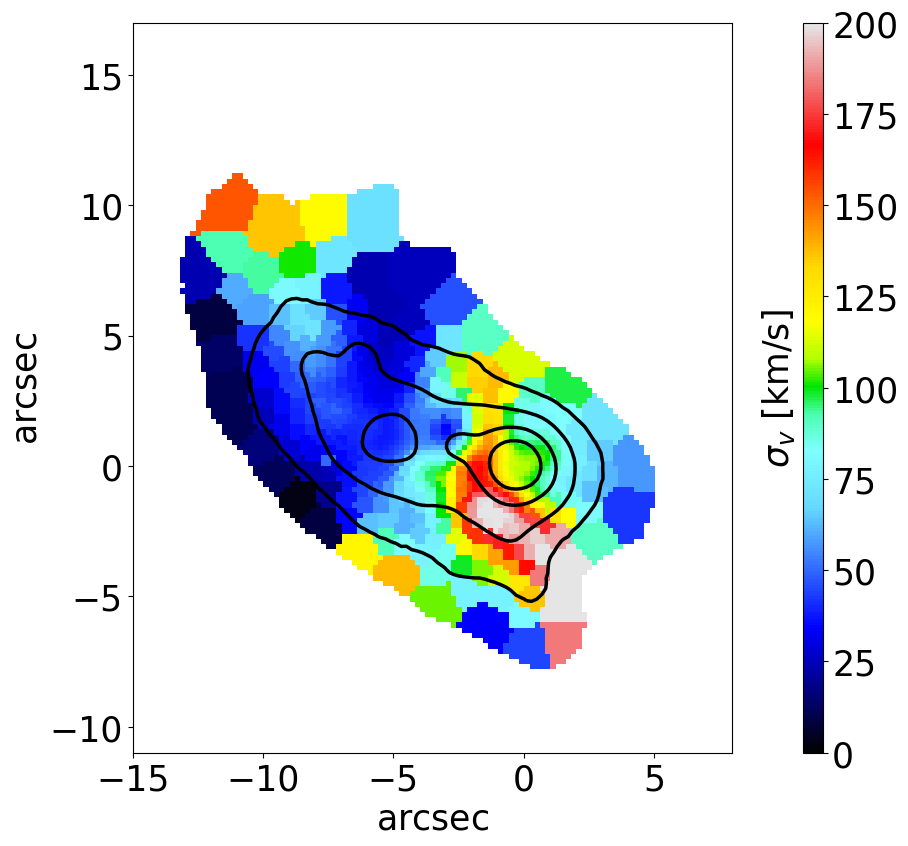}
\caption{Stellar kinematics maps for Voronoi bins with S/N> 10.  Left:  velocity. Right:  velocity dispersion.   Contours are continuum isophotes as in Fig.\ref{fig:white}. \label{fig:vel_star}}
\end{figure*}

The left panel of Figure \ref{fig:vel_star} shows the velocity field of the stellar component. We choose as zero point of the velocity the same of that of the gas.   Two distinct constituents are visible: the East side of the galaxy presents low stellar velocities, ranging from $\sim -80 \, \rm km \, s^{-1}$ to $\sim 80 \,  \rm km \, s^{-1}$. In this region  stars and gas have similar velocities. In contrast, the West side of the galaxy presents much higher velocities, in the range $200<v (\rm km \, s^{-1}) <350$. 
The median error on the velocities is 36 km s$^{-1}$. 
The relative velocity of the two sides of the system is  $\sim 200 \, \rm km \, s^{-1}$, suggesting that they are indeed physically associated.

The velocity dispersion map shown in the right panel of Figure \ref{fig:vel_star} is very chaotic and regular trends are not detectable. The North-East side generally shows lower $\sigma_{star}$ values ($\sim 30 \, \rm km \,s^{-1}$), while the South-West side is characterized by much higher values (up to $\sim 250 \rm \, km\, s^{-1}$). 
The median error on the velocity dispersions is 57 km s$^{-1}$.

\subsection{Spatially resolved stellar properties}

\begin{figure}
\centering
\includegraphics[scale=0.36]{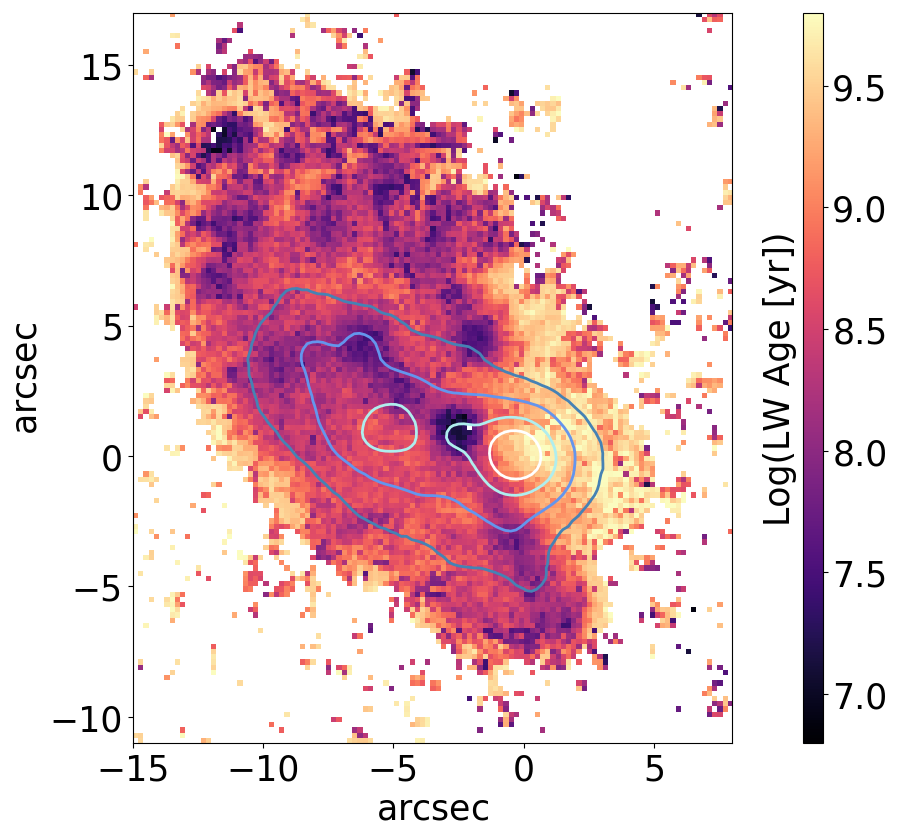}
\caption{Luminosity weighted age (LWA) map. Contours are continuum isophotes as in Fig.\ref{fig:white}. \label{fig:LWA}}
\end{figure}

The distribution of the stellar luminosity weighted age presented in Fig. \ref{fig:LWA} follows a similar spatial pattern as the one depicted in the stellar kinematics. 
The region Western to the ($x\sim0$) position has a typical age of $5\times 10^9$ yr, while the rest of the galaxy is characterized by much younger ages, of $5\times 10^8$ yr. The knot A 
has a even younger age, $\sim 10^7$ yr. The median relative error on the luminosity weighted ages is of the order of  15\%.

\begin{figure*}
\centering
\includegraphics[scale=0.36]{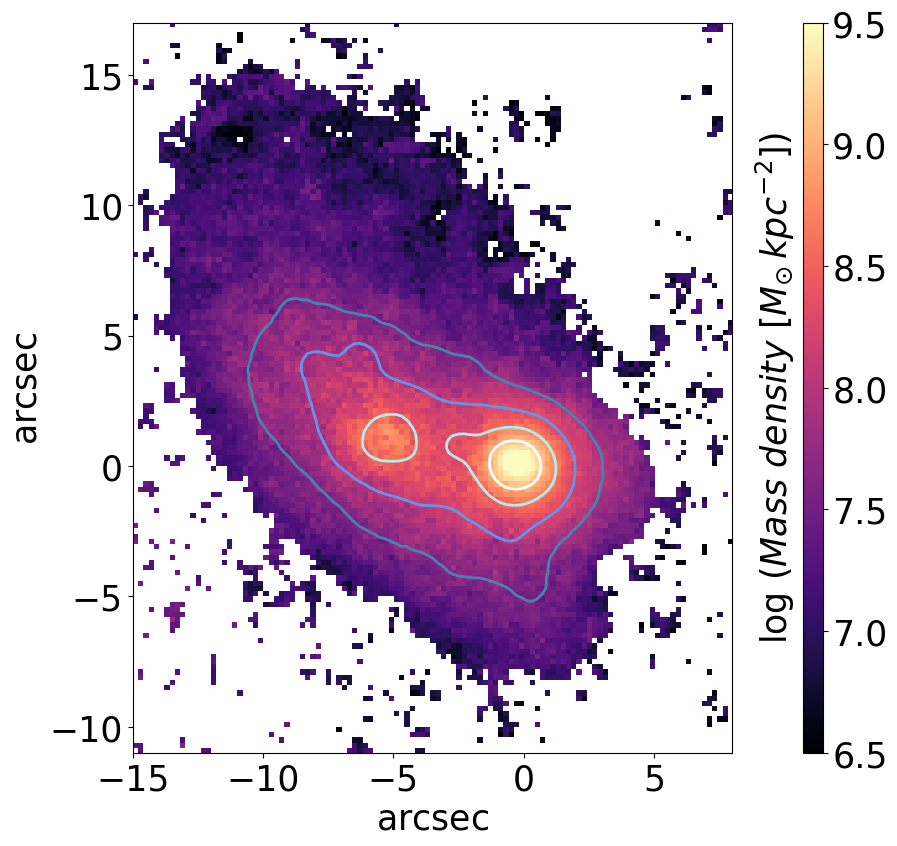}
\includegraphics[scale=0.36]{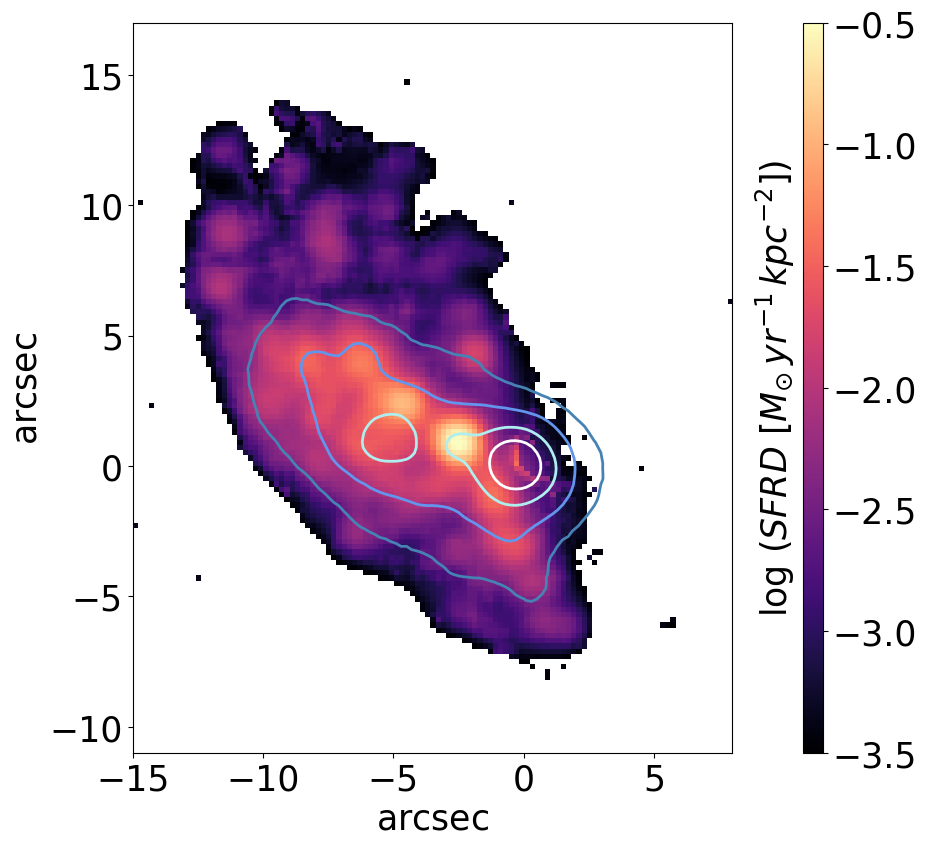}
\caption{{\it Left:} Stellar mass density map. {\it Right:} Star formation density map. In both panels, contours are continuum isophotes as in Fig.\ref{fig:white}. \label{fig:sfr_mass}}
\end{figure*}

The stellar mass density shown in the left panel of Fig. \ref{fig:sfr_mass} is characterized by a double peak, roughly corresponding to the peaks of the stellar continuum. 
The brighter peak has a stellar mass density as high as $\rm 5\times 10^9 M_\odot/kpc^2$, while the secondary peak, which is also more spread, reaches  values of $\rm 5\times 10^8 M_\odot/kpc^2$. The Northern part of the galaxy carries much less mass: its typical surface mass density is  $\rm \sim  5\times 10^7 M_\odot/kpc^2$. 
Running {\sc sinopsis} on the integrated spectra of the entire galaxy, we obtain a total M$_\ast$ of 2.5$\rm \times 10^{10} \, M_\odot$,

The mass density does not trace the  distribution of the recent star formation, shown in the right panel of   Fig. \ref{fig:sfr_mass}.  The median error on the logarithm of SFRD,  computed propagating the error on the \Ha flux as computed by {\sc kubeviz}, is 0.01 dex.
Indeed, there is almost no  star formation at the position of the densest component. The peak of the star formation is taking place in the knot A.
An arc-shape trail of highly star forming regions centered  at this position extends both toward South-West and North-East.  Star forming regions are also  evident on the northernmost edge of the galaxy.
Integrating the spectrum over all the spaxels with S/N(\Ha)$>4$, we get a value of SFR=1.8 $\rm{M_\odot \, yr^{-1}}$. 

\begin{figure*}
\centering
	\includegraphics[scale=0.36]{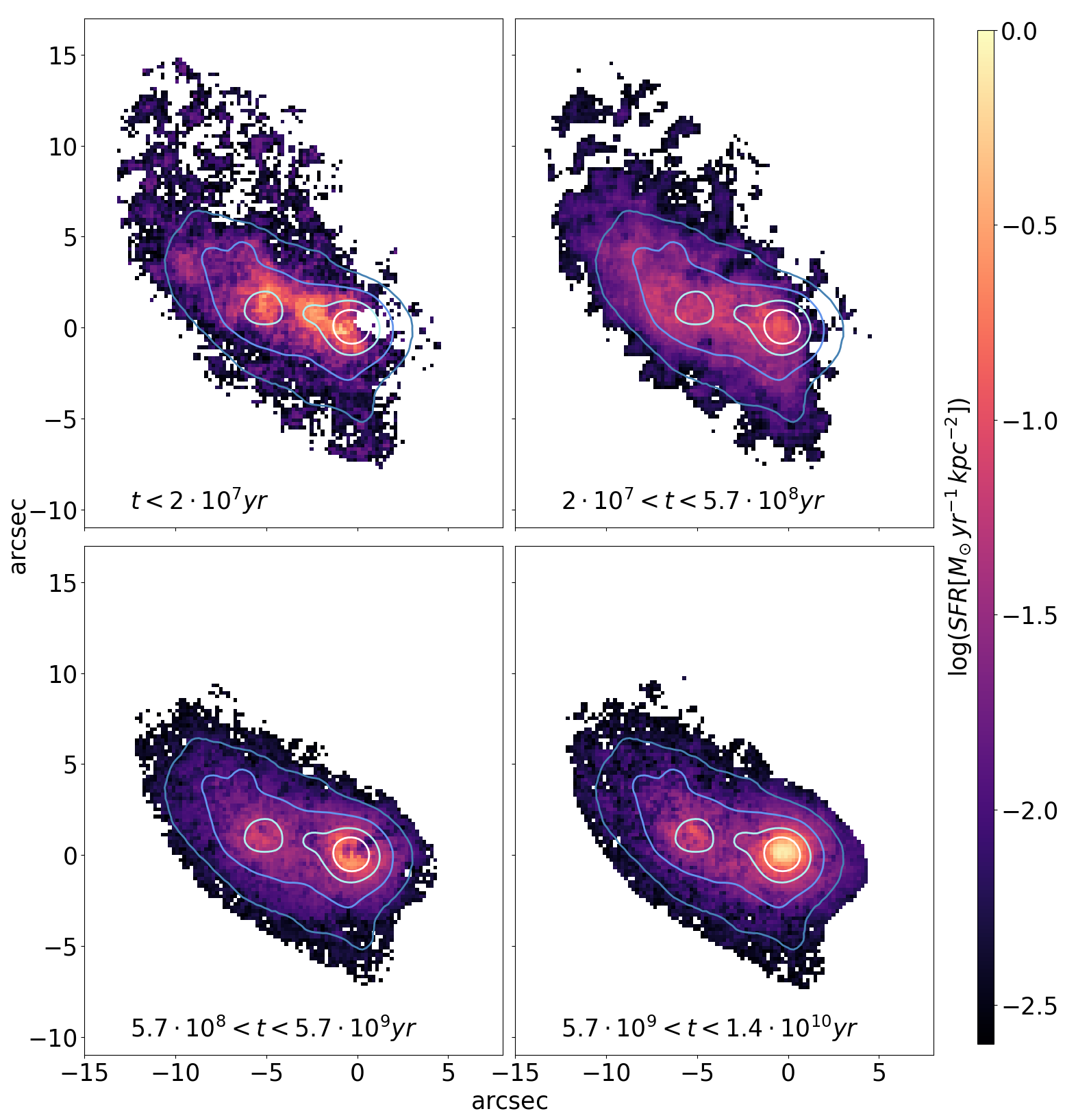}
\caption{Average star formation rate per kpc$^2$ in four different ages: 
during the last $2\times 10^7$ yr (top left), between $2\times 10^7$yr and $5.7 \times 10^8$yr (top right), $5.7 \times 10^8$yr and $5.7 \times10^9$yr (bottom left) and $> 5.7 \times 10^9$yr ago (bottom right). Contours are continuum isophotes as in Fig.\ref{fig:white}. \label{fig:SFH_maps}}
\end{figure*}

 Figure \ref{fig:SFH_maps} shows the  spatially resolved star formation history. This map allows us to characterize how many stars were formed at each spatial location as a function of cosmic time.  We choose  four logarithmically spaced age bins in such a way that the differences between the spectral characteristics of the stellar populations are maximal (\citealt{Fritz2007} and \citetalias{Fritz2017}). The median errors on the SFR per kpc$^2$ are of the order of 0.003 $M_\odot \, yr^{-1} \, kpc^{-2}$ in the two youngest age bins, of the order of 0.007 $M_\odot \, yr^{-1} \, kpc^{-2}$ in the two oldest age bins.
 The upper left panel shows that the ongoing SFR ($t<2\times 10^7$ yr)
 is very intense in the central region of the system, but is essentially absent in the Westernmost side where there is no ionized gas (Fig.\ref{fig:Ha}). High star formation activity is observed in the knot A, while low-levels of star formation are detected in the Northern part of the galaxy. 
The recent star formation activity ($2\times 10^7$ yr$<t<5.7\times 10^8$ yr, upper right panel) has a similar spatial distribution compared to the youngest stars, even though recent star formation extends towards the western part of the galaxy, while the northern part is less extended. At these ages, the knot A is not visible anymore. 

As we move to the intermediate ($5.7\times 10^8$ yr$<t<5.7\times 10^9$ yr, bottom left panel) and  old ($t>5.7\times 10^9$ yr, bottom right panel) populations, the distribution of the stars  is different, being mostly confined within the continuum isophotes. All the westernmost region of the galaxy was star forming at those epochs, as was the  brightest of the two components of the system.

\subsection{Properties of the two merging components}

\begin{figure}
\centering
\includegraphics[scale=0.26]{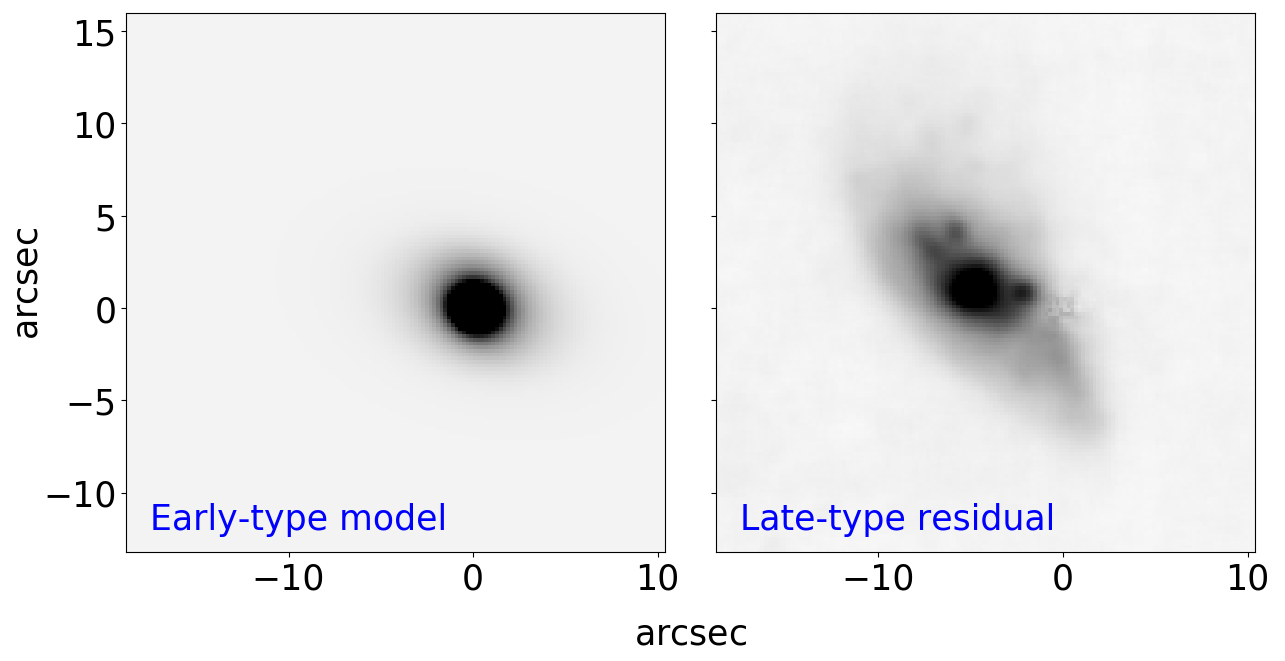}
\caption{{\it Left.} Model for the early-type component based on the MUSE white-light image.  {\it Right.} Residual image for the late-type galaxy. See text for details. \label{fig:model}}
\end{figure}

To characterize the morphology of the two objects of the analyzed system, we use the white-light image shown in Fig.\ref{fig:white} to model the two components, using an iterative method to separate their contribution to the total light. In particular, the purposely devised software AIAP (by Giovanni Fasano, private communication) allows us to perform adaptive smoothing, interactive masking and ellipse fitting of each galaxy isophote, as well as to generate a smooth model faithfully reproducing the shape of the real galaxies, including position angle and ellipticity profiles. Such an accurate modeling procedure has been iteratively applied to the early-type-like (first step) and late-type components of the system, each time removing the model from the original image. The end products of this iteration are the final  model of the early-type-like galaxy and the residual image of the late-type galaxy (see Fig. \ref{fig:model}).

Fitting the early-type component with a Sersic law, we obtian a value for the Sersic index $n$ of 2.51 and a value for  the effective radius $R_e$ of 1.8 kpc.  
The total luminosity in the wavelength range 4750-9350 \AA{}  is 
2.2$\rm \times 10^{19} \, erg/s$.
The parameters of the late-component are instead $n$=0.2, $R_e$=5.3 kpc, 
 total luminosity =2.9$\rm \times 10^{19} \, erg/s$. 

The median LWA for the early-type galaxy is $\sim 2\times 10^9$yr,  that for the late-type galaxy is instead significantly lower: $\sim 2.5\times 10^8$yr.

Running {\sc sinopsis} on the integrated spectra of the early type (obtained from the aforementioned model),  
we obtain a $M_\ast$ of 1.4$\rm \times 10^{10} \, M_\odot$. Subtracting this value to the total stellar mass of the system, we infer a stellar mass for the late-type component equal to $\rm \sim 1.1\times 10^{10} \, M_\odot$.
 These values have to be taken with caution, given the physical spatial overlap of the two galaxies, with the consequence that a number of spaxels contribute to both components.

\section{Discussion}
In the previous section we have described the most salient  properties of \pp. Here we connect all of them to draw a scenario able to  explain the formation and evolution of the galaxy.

\pp is part of a small group with three members. Nonetheless, we have not identified any sign of disturbance due to the presence of the other two galaxies, which are located at a projected distance of 540 kpc and 750 kpc, respectively.

Our interpretation is that we are witnessing an ongoing $\sim$1:1 merger between two galaxies with distinct properties. 

A gas-poor early-type galaxy has probably recently merged with a late-type one. 
The relative motion of the stars of the two components is $\sim 200 \rm \, km \, s^{-1}$, indicating that the two systems are indeed merging. Currently, the early-type galaxy has a higher line-of-sight velocity. 
The gas motion is disturbed on the side of the collision. 

The absence of prominent tidal tails, which usually form after the first passage \citep[e.g.][]{Mihos1996}, in the late-type indicates that the early-type is in its first approach. Note, however, that tails often fade away at large distances due to surface brightness dimming \citep[e.g.][]{Hibbard1997, Overzier2010, Hung2014}. 

The early-stage of the merger is also  corroborated by the rather strong metallicity gradient we have observed in the system. 
Indeed, models of inside-out disk growth predict that gradients should be initially steep and become flatter at later times \citep[e.g. ][]{Prantzos2000, Magrini2007, Fu2009, Marcon2010}, after  the gas inflows of relatively metal-poor gas 
has had the time to  change the galaxy's metal distribution \citep{Mihos1996, Barnes1996}. N-body/SPH numerical simulations of equal-mass mergers by \cite{Rupke2010} also show that between first and second pericenter the initial radial metallicity profile noticeably flattens. This smoothing reflects the effects of gas redistribution over the galaxy disk.

During the interaction, there has been an enhancement of the star formation activity in the central part of the system. \cite{Moreno2015}, employing  a suite of 75 SPH merger simulations, have found that whenever enhanced star formation in the nucleus is triggered, this is always accompanied by the suppression of the star formation activity at large galactocentric radii, similarly to what we observe in \pp. 

The high star formation gives rise to enhancements in the  metallicity of the central part of the galaxy, in agreement to predictions of N-body/smooth particle hydrodynamics models \citep{Torrey2012b}. 

The overall gas redistribution  does not excessively boost the star formation: \pp  lays on the typical SFR-mass relation for star-forming field galaxies  \citep[see][]{Poggianti2016}. This is true both if we consider the  total stellar mass and if we 
assume that all the star formation comes from the late-type galaxy. 

\begin{figure*}
\centering
\includegraphics[scale=0.38]{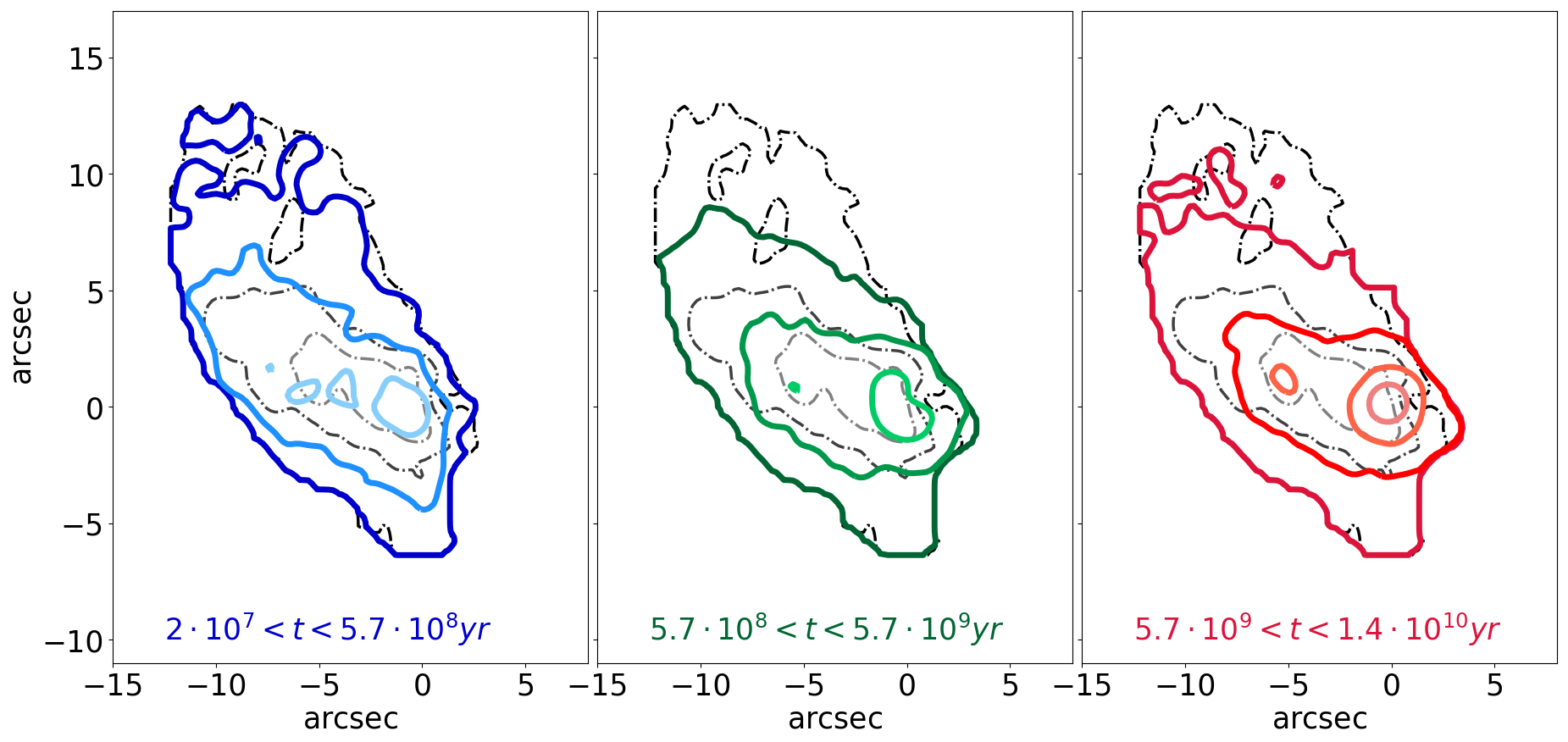}
\caption{Comparison between the stellar maps of different ages. The ongoing star formation bin ($t<2\times 10^7$ yr, black and gray lines) is shown in all the panels, for reference. {\it Left:} recent star formation bin ($\rm 2\times 10^7  yr <t<5.7\times 10^8  yr$, blue lines). {\it Middle:} intermediate star formation bin ($\rm 5.7\times 10^8 yr <t<5.7\times 10^9  yr$, green lines). {\it Right:} old star formation bin ($t> 5.7 \times 10^9$yr, red lines).  Contours are logarithmically spaced between SFR= 0.0001 and 0.1 $M_\odot \, yr^{-1}$.\label{fig:cfr_pops}}
\end{figure*}

We can try to date the merging event comparing the stellar maps of different ages, as done in Fig.\ref{fig:cfr_pops}. The contours representing the populations with $t> 5.7 \times 10^9$yr highlight the presence of two distinct peaks in the star formation distribution, with less stars forming between them. On the contrary, the populations with $t<5.7\times 10^8  yr$ highlight the presence of new born material in between the two components, suggesting that the interaction has started $\rm 2\times 10^7  yr <t<5.7\times 10^8  yr$ ago and is still ongoing. The forming stellar structure in the central part of the merging system corresponds to the region that we have called knot A. 
Approximately at the same time, the merger has been also moving gas towards the Norhtern part of the galaxy, where new-born stars are observed. New high resolution simulations \citep{Teyssier2010, Hopkins2013, Renaud2015} show that during the first two pericenter passages extended star formation arises spontaneously. This is most likely due to fragmentation of the gas clouds produced by an increase of the interstellar medium  supersonic turbulence as a consequence of the tidal interaction itself \citep{Renaud2014}. 

The build up of new stars  produces a lopsided morphology to the resulting galaxy. As also  shown in Fig.\ref{fig:SFH_maps}, the gas rich object had a  regular shape at the early stages of its formation, while for $t<5.7\times 10^8  yr$ it started developing an asymmetric shape. 

Lopsidedness induced by tidal encounter between two galaxies with an arbitrary orientation is a common feature in group galaxies \citep[e.g.][]{Combes2004}, where  the perturbation is expected to  generate a force term, responsible for the  lopsidedness in the galaxy. Lopsidedness can also be generated due to a mild interaction or more indirectly due to the response of the disk to the distorted halo which feels a stronger effect of the interaction \citep{Weinberg1995, Jog1997, Schoenmakers1997}.

 \subsection{The formation of a Tidal Dwarf Galaxy}\label{sec:tdg}
 As mentioned in the previous section, the merger between the two galaxies induces the formation of a number of   stellar structures. 
The conditions in which the structures form determine whether it is most likely SSCs or TDGs (see Sect.\ref{sec:intro}). 

According to \cite{Duc2012}, TDGs should mostly be produced by major wet mergers that occurred less than one Gyr ago with low impact velocities (up to 250 $\rm km \, s^{-1}$) involving at least one spiral galaxy and are formed by material that used to belong to the parent galaxies.  The production rate is  about 1 TDG per favorable merger \citep[see also][]{Bournaud2007}. 
 Being recycled objects, TDGs have inherited the metal content of the interstellar medium from their progenitors. Thus, their metallicity gives information about the past chemical enrichment of the progenitors, and is  not correlated with their current mass, contrary to normal galaxies. 
TDGs have masses around $10^8\, M_\odot$, and are young objects. They still exhibit the tails and bridges in which they were formed that have not had the time to evaporate yet. Once evolved, TDGs should become indistinguishable from regular satellite galaxies on optical images.

According to simulations \citep{Duc2004, Bournaud2006, Wetzstein2007}, TDGs should also be free of dark matter. Indeed,   their low escape velocity is low (at most a few tens of $\rm km \,  s^{-1}$), therefore the dark matter particles of the halo of the parent spiral galaxy  will not be held by the gravitational well of the TDG. 

Since TDGs  are made-up  from material coming from the disk of the progenitor, the dynamical mass measured from their rotation velocity and size should be similar to their visible mass in stars and gas \citep{Zackrisson2010,Bournaud2010}.
 
In our analysis we have have paid particular attention to a structure formed from the impact between the two galaxies, located at ($x=-2.5^{\prime\prime}$, $y= 1^{\prime\prime}$) that we called knot A.

The location and radius of the knot are found through a purposely devised script developed in the context of the GASP survey \citepalias[see][]{Poggianti2017a}. The knot's radius is estimated through a recursive analysis, taking care of deblending the contributions from overlapping sources   
and is $\sim 2.2$ kpc. This value is well above the seeing ($\sim 1^{\prime\prime}$) and the MUSE PSF. 

We then run {\sc kubeviz} and {\sc sinopsis} on a mask identifying only the knot, to obtain its properties. As the analysis of emission-line ratios shows that star formation is the main source of ionization, we can also compute the electron density $n_e$ and the  mass of the ionized gas $M_{gas}$, following the relation presented in \cite{Proxauf2014} and described in detail in \citetalias{Poggianti2017a}.

We  estimate the following quantities for the knot A:  $A_V$=0.85 mag, $12 + \log(O/H)$ =8.4, $n_e=54cm^{-3}$, $ M_{gas}=2.5\times10^6 M_\odot$,  $ M_{star}=6.3\times 10^9 M_\odot$, mass density=$4.1\times 10^8 M_\odot \, kpc^{-2}$,  luminosity weighted age =$\sim 3\times 10^{7}$ yr, SFR=0.3$M_\odot\, yr^{-1}$, SFRD=0.02$M_\odot\, yr^{-1}\, kpc^{-2}$. 

This knot therefore is very massive and large, has one of the largest metallicity values of the system, and is relatively young. It has also a  high SFR and it is relatively dusty.  It was born in correspondence of the merger, and  connection with the parent galaxy is still visible and is most likely a spiral arm. 
Its dynamical mass is $M_{dyn}=2\times r\frac{\sigma_r^2}{G}\sim 2.5 \times 10^9 M_\odot$ if we adopt 
 $R_e\sim 2.2 kpc$
and  $\sigma_r\sim $50 $\rm km \, s^{-1}$. This value is very similar to its stellar mass, suggesting that there is not dark matter in the structure.  

This object is therefore a good candidate for being or soon becoming a TDG. 
Nonetheless, it presents peculiar properties. 

\cite{Kaviraj2012} have conducted a statistical observational study of the TDGs in the nearby Universe. They found that 95\% of TDG-producing mergers involve interactions between two blue, spiral galaxies. The vast majority of these parent systems have mass ratios greater than $\sim$ 1:7. Only a fifth of gas-rich major mergers produce TDGs with masses greater than $10^8 M_\odot$. TDGs are usually within $\sim$20 kpc of their parent galaxies.

Differently from other TDG candidates \citep[see also][]{Sengupta2014}, knot A  has formed in a very early stage of the merger and even though it is still in the phase of forming and still located within the disk of its progenitor,  it has already been able to accrete great quantities of gas and stars and has high and unusual SFR values. 

Another explanation might be that instead of being a merger-induced TDG, the observed structure is a huge (up to a few $10^8 M_\odot$) star-forming region, similarly to what detected at similar redshift by e.g. \cite{Fisher2014} and at higher redshift ($1<z<3$) by e.g. \cite{Genzel2011, Elmegreen2005, Elmegreen2009}. However, this interpretation requires disk-dominated systems and high gas fractions (>30-50\%), which are not the case for \pp. We can roughly estimate 
the gas fractions from the stellar mass and star formation density maps shown in Fig.\ref{fig:sfr_mass} and the \cite{Kennicutt1998b} law. We get a gas fraction of $\sim$ 10-15\% if the system is on the low-efficiency side on the relation, and even less if the system is on the higher side, as expected for a merger \citep{Daddi2010}. As a consequence, both the TDG and the other structures detected as bright blue regions in formation (Fig.\ref{fig:rgb_image}) are most likely merger-induced and not simply  large SF regions intrinsic to this particular system.

A third option might be that knot A is a tidally induced star forming region that was triggered by the tidal/compressive forces of the merger. This possibility  would not require large disk gas fractions. If the structure will remain kinematically bound to the parent galaxy, it will not turn into a TDG. 
As we are observing knot A in an early stage of its life, we can not firmly state that it will certainly become a long-lived TDG. However, its peculiar properties, and in particular its very high velocity dispersion makes it  a remarkable object and a very good TDG candidate, rather than simply a massive tidally-induced star forming region within a merging system.

To conclude, we note that knot A seems to be in a filament of star formation regions triggered by the merger. Indeed, any merger is expected to create a number of star-forming knots,  with a range of masses. However, simulations show that only the most massive ones tend to survive \citep[see, e.g.][]{Bournaud2006}.
We can identify two other structures with quite peculiar characteristics in the filament. These are located at ($x=-5^{\prime\prime}$,  $y=3^{\prime\prime}$), and at ($x=-7^{\prime\prime}$, $y=5^{\prime\prime}$), respectively. In addition to be smaller in size (radius $\sim 1.6$ and 1.8 kpc, respectively), they are also considerably less star forming (SFR$\rm \sim0.09 M_\odot \, yr^{-1}$, SFRD$\rm \sim0.001$ and $0.009 M_\odot \, yr^{-1}\, kpc^{-2}$), less massive ($\rm M_\ast\sim 8\times10^8$ and  $\rm M_\ast\sim 1.4\times10^9 M_\odot$, mass density$\sim 10^8$ and  $\rm M_\ast\sim 1.4\times10^8 M_\odot\, kpc^{-2}$), older (luminosity weighted age $\sim 1.3\times 10^8$ and $\sim 1.5\times 10^8 yr$) and are characterised by lower values of velocity dispersion ($\rm \sim 25 km\, s^{-1}$). Rather than being TDG candidates, these will therefore most likely become SSCs, or, if massive enough, eventually globular clusters \citep{Bournaud2008}.

\section{Conclusions}
GASP (GAs Stripping phenomena in galaxies with MUSE) is an ongoing ESO Large Program with the MUSE/VLT to investigate the causes and the effects of gas removal processes in galaxies in different environments. The sample includes  galaxies  selected for showing signs that could be indicative of stripping in B-band images.

Within the sample, we identified an ongoing merger between two galaxies with different properties.  This peculiar system is so far unique in the GASP sample, suggesting that usually mergers are not confused with galaxies undergoing  gas stripping.

A gas poor,  early-type-like object ($n=2.51$, $R_e= 1.8$ kpc, $M_\ast=1.4 \times 10^{10} M_\odot$) with  a luminosity weighted age of $\rm \sim 2\times 10^9 \, yr$  has started an interaction with a  younger (luminosity weighted age $\rm \sim 2.5\times 10^8 \, yr$), gas rich,  late-type object ($n=0.2$, $R_e= 5.3$ kpc, $M_\ast= 1.1\times 10^{10} M_\odot$) between $\rm 2\times 10^7  yr <t<5.7\times 10^8  yr$ ago. The merger  is still in an early phase.  The gas kinematic pattern reflects the gas of the late-type object and is mainly distorted in correspondence to the location of the impact, while the northern regions had not time yet to be influenced. The stellar kinematic instead is much more chaotic, as expected in case of mergers. Nonetheless, the rotation of the stars in the two galaxies is still detectable.  The early-type galaxy has currently a higher line-of-sight velocity than the late-type galaxy.

The merger induces a gas redistribution  igniting high levels of star formation both in the northern part of the galaxy, producing a lopsidedness visible for $t<5.7\times 10^8 \, yr$, and between the two galaxies, especially in the region of the impact. As a consequence, a large stellar structure forms. This object presents peculiar properties that make it a good TDG candidate. Indeed, it  has formed in a very early stage of the merger and even though it is still in the phase of forming and still located within the disk of its progenitor,  it has already  accreted great quantities of gas and stars ($M_\ast \sim 6\times 10^9 M_\odot$, $R\sim 2.2$ kpc) and has high and unusual SFR values (SFR$\sim 0.3 M_\odot \, yr^{-1}$). 

This is the first detailed characterization of a TDG candidate detected beyond the local Universe, where data have relatively lower resolution. It could pave the way to search for TDGs at higher redshift, a regime still poorly characterized.

\acknowledgments
We thank the referee for their useful comments that helped us to improve the manuscript. Based on observations collected at the European Organisation for Astronomical Research in the Southern Hemisphere under ESO programme 196.B-0578. 
This work made use of the {\sc kubeviz} software which is publicly available at \url{http://www.mpe.mpg.de/~dwilman/kubeviz/}. 
We are grateful to Joe Liske, Simon Driver and the whole MGC collaboration for making their dataset easily available, and to Rosa Calvi for her valuable work on the PM2GC. We also thank Fredric Bournaud for helpful comments on a draft of this paper.
We acknowledge financial support from PRIN-INAF 2014. B.V. acknowledges the support from 
an  Australian Research Council Discovery Early Career Researcher Award (PD0028506).  J.F. acknowledges financial support from UNAM-DGAPA-PAPIIT IA104015 grant, M\'exico.
This work was co-funded under the Marie Curie Actions of the European Commission (FP7-COFUND). 

\facilities{VLT(MUSE)} 
\software{KUBEVIZ, ESOREX, SINOPSIS, IRAF, CLOUDY, pyqz, IDL, Python}

\bibliographystyle{yahapj}
\bibliography{gasp}

\begin{thebibliography}{}
\providecommand\natexlab[1]{#1}
\providecommand\JournalTitle[1]{#1}

\bibitem[{{Ahn} {et~al.}(2012){Ahn}, {Alexandroff}, {Allende Prieto},
  {Anderson}, {Anderton}, {Andrews}, {Aubourg}, {Bailey}, {Balbinot}, {Barnes},
  \& et~al.}]{Ahn2012}
{Ahn}, C.~P., {Alexandroff}, R., {Allende Prieto}, C., {et~al.} 2012,
  \href{http://dx.doi.org/10.1088/0067-0049/203/2/21}{\JournalTitle{\apjs},
  203, 21}

\bibitem[{{Baldwin} {et~al.}(1981){Baldwin}, {Phillips}, \&
  {Terlevich}}]{Baldwin1981}
{Baldwin}, J.~A., {Phillips}, M.~M., \& {Terlevich}, R. 1981,
  \href{http://dx.doi.org/10.1086/130766}{\JournalTitle{\pasp}, 93, 5}

\bibitem[{{Barnes} \& {Hernquist}(1996)}]{Barnes1996}
{Barnes}, J.~E., \& {Hernquist}, L. 1996,
  \href{http://dx.doi.org/10.1086/177957}{\JournalTitle{\apj}, 471, 115}

\bibitem[{{Barnes} \& {Hernquist}(1991)}]{Barnes1991}
{Barnes}, J.~E., \& {Hernquist}, L.~E. 1991,
  \href{http://dx.doi.org/10.1086/185978}{\JournalTitle{\apjl}, 370, L65}

\bibitem[{{Blumenthal} {et~al.}(1984){Blumenthal}, {Faber}, {Primack}, \&
  {Rees}}]{Blumenthal1984}
{Blumenthal}, G.~R., {Faber}, S.~M., {Primack}, J.~R., \& {Rees}, M.~J. 1984,
  \href{http://dx.doi.org/10.1038/311517a0}{\JournalTitle{\nat}, 311, 517}

\bibitem[{{Boquien} {et~al.}(2007){Boquien}, {Duc}, {Braine}, {Brinks},
  {Lisenfeld}, \& {Charmandaris}}]{Boquien2007}
{Boquien}, M., {Duc}, P.-A., {Braine}, J., {et~al.} 2007,
  \href{http://dx.doi.org/10.1051/0004-6361:20066692}{\JournalTitle{\aap}, 467,
  93}

\bibitem[{{Bournaud}(2010)}]{Bournaud2010}
{Bournaud}, F. 2010, in Astronomical Society of the Pacific Conference Series,
  Vol. 423, Galaxy Wars: Stellar Populations and Star Formation in Interacting
  Galaxies, ed. B.~{Smith}, J.~{Higdon}, S.~{Higdon}, \& N.~{Bastian}, 177

\bibitem[{{Bournaud} \& {Duc}(2006)}]{Bournaud2006}
{Bournaud}, F., \& {Duc}, P.-A. 2006,
  \href{http://dx.doi.org/10.1051/0004-6361:20065248}{\JournalTitle{\aap}, 456,
  481}

\bibitem[{{Bournaud} {et~al.}(2008){Bournaud}, {Duc}, \&
  {Emsellem}}]{Bournaud2008}
{Bournaud}, F., {Duc}, P.-A., \& {Emsellem}, E. 2008,
  \href{http://dx.doi.org/10.1111/j.1745-3933.2008.00511.x}{\JournalTitle{\mnras},
  389, L8}

\bibitem[{{Bournaud} {et~al.}(2007){Bournaud}, {Duc}, {Brinks}, {Boquien},
  {Amram}, {Lisenfeld}, {Koribalski}, {Walter}, \&
  {Charmandaris}}]{Bournaud2007}
{Bournaud}, F., {Duc}, P.-A., {Brinks}, E., {et~al.} 2007,
  \href{http://dx.doi.org/10.1126/science.1142114}{\JournalTitle{Science}, 316,
  1166}

\bibitem[{{Brinchmann} {et~al.}(2004){Brinchmann}, {Charlot}, {White},
  {Tremonti}, {Kauffmann}, {Heckman}, \& {Brinkmann}}]{Brinchmann2004}
{Brinchmann}, J., {Charlot}, S., {White}, S.~D.~M., {et~al.} 2004,
  \href{http://dx.doi.org/10.1111/j.1365-2966.2004.07881.x}{\JournalTitle{\mnras},
  351, 1151}

\bibitem[{{Bundy} {et~al.}(2015){Bundy}, {Bershady}, {Law}, {Yan}, {Drory},
  {MacDonald}, {Wake}, {Cherinka}, {S{\'a}nchez-Gallego}, {Weijmans}, {Thomas},
  {Tremonti}, {Masters}, {Coccato}, {Diamond-Stanic}, {Arag{\'o}n-Salamanca},
  {Avila-Reese}, {Badenes}, {Falc{\'o}n-Barroso}, {Belfiore}, {Bizyaev},
  {Blanc}, {Bland-Hawthorn}, {Blanton}, {Brownstein}, {Byler}, {Cappellari},
  {Conroy}, {Dutton}, {Emsellem}, {Etherington}, {Frinchaboy}, {Fu}, {Gunn},
  {Harding}, {Johnston}, {Kauffmann}, {Kinemuchi}, {Klaene}, {Knapen},
  {Leauthaud}, {Li}, {Lin}, {Maiolino}, {Malanushenko}, {Malanushenko}, {Mao},
  {Maraston}, {McDermid}, {Merrifield}, {Nichol}, {Oravetz}, {Pan}, {Parejko},
  {Sanchez}, {Schlegel}, {Simmons}, {Steele}, {Steinmetz}, {Thanjavur},
  {Thompson}, {Tinker}, {van den Bosch}, {Westfall}, {Wilkinson}, {Wright},
  {Xiao}, \& {Zhang}}]{Bundy2015}
{Bundy}, K., {Bershady}, M.~A., {Law}, D.~R., {et~al.} 2015,
  \href{http://dx.doi.org/10.1088/0004-637X/798/1/7}{\JournalTitle{\apj}, 798,
  7}

\bibitem[{{Calvi} {et~al.}(2011){Calvi}, {Poggianti}, \& {Vulcani}}]{Calvi2011}
{Calvi}, R., {Poggianti}, B.~M., \& {Vulcani}, B. 2011,
  \href{http://dx.doi.org/10.1111/j.1365-2966.2011.19088.x}{\JournalTitle{\mnras},
  416, 727}

\bibitem[{{Cappellari}(2012)}]{Cappellari2012}
{Cappellari}, M. 2012, {pPXF: Penalized Pixel-Fitting stellar kinematics
  extraction}, Astrophysics Source Code Library,
  \href{http://arxiv.org/abs/1210.002}{{\sffamily ascl:1210.002}}

\bibitem[{{Cappellari} \& {Copin}(2012)}]{Cappellari2012_v}
{Cappellari}, M., \& {Copin}, Y. 2012, {Voronoi binning method}, Astrophysics
  Source Code Library, \href{http://arxiv.org/abs/1211.006}{{\sffamily
  ascl:1211.006}}

\bibitem[{{Cardelli} {et~al.}(1989){Cardelli}, {Clayton}, \&
  {Mathis}}]{Cardelli1989}
{Cardelli}, J.~A., {Clayton}, G.~C., \& {Mathis}, J.~S. 1989,
  \href{http://dx.doi.org/10.1086/167900}{\JournalTitle{\apj}, 345, 245}

\bibitem[{{Chabrier}(2003)}]{Chabrier2003}
{Chabrier}, G. 2003,
  \href{http://dx.doi.org/10.1086/376392}{\JournalTitle{\pasp}, 115, 763}

\bibitem[{{Chapon} {et~al.}(2013){Chapon}, {Mayer}, \& {Teyssier}}]{Chapon2013}
{Chapon}, D., {Mayer}, L., \& {Teyssier}, R. 2013,
  \href{http://dx.doi.org/10.1093/mnras/sts568}{\JournalTitle{\mnras}, 429,
  3114}

\bibitem[{{Combes}(2003)}]{Combes2003}
{Combes}, F. 2003, in Astronomical Society of the Pacific Conference Series,
  Vol. 290, Active Galactic Nuclei: From Central Engine to Host Galaxy, ed.
  S.~{Collin}, F.~{Combes}, \& I.~{Shlosman}, 411

\bibitem[{{Combes} {et~al.}(2004){Combes}, {Boisse}, {Mazure}, \&
  {Blanchard}}]{Combes2004}
{Combes}, F., {Boisse}, P., {Mazure}, A., \& {Blanchard}, A. 2004

\bibitem[{{Cortijo-Ferrero} {et~al.}(2017{\natexlab{a}}){Cortijo-Ferrero},
  {Gonz{\'a}lez Delgado}, {P{\'e}rez}, {Cid Fernandes}, {S{\'a}nchez}, {de
  Amorim}, {Di Matteo}, {Garc{\'{\i}}a-Benito}, {Lacerda}, {L{\'o}pez
  Fern{\'a}ndez}, \& {Tadhunter}}]{Cortijo2017a}
{Cortijo-Ferrero}, C., {Gonz{\'a}lez Delgado}, R.~M., {P{\'e}rez}, E., {et~al.}
  2017{\natexlab{a}},
  \href{http://dx.doi.org/10.1093/mnras/stx383}{\JournalTitle{\mnras}, 467,
  3898}

\bibitem[{{Cortijo-Ferrero} {et~al.}(2017{\natexlab{b}}){Cortijo-Ferrero},
  {Gonz{\'a}lez Delgado}, {P{\'e}rez}, {Cid Fernandes}, {Garc{\'{\i}}a-Benito},
  {Di Matteo}, {S{\'a}nchez}, {de Amorim}, {Lacerda}, {L{\'o}pez
  Fern{\'a}ndez}, \& {Tadhunter}}]{Cortijo2017c}
---. 2017{\natexlab{b}}, \JournalTitle{ArXiv e-prints},
  \href{http://arxiv.org/abs/1707.05324}{{\sffamily arXiv:1707.05324}}

\bibitem[{{Cortijo-Ferrero} {et~al.}(2017{\natexlab{c}}){Cortijo-Ferrero},
  {Gonz{\'a}lez Delgado}, {P{\'e}rez}, {S{\'a}nchez}, {Cid Fernandes}, {de
  Amorim}, {Di Matteo}, {Garc{\'{\i}}a-Benito}, {Lacerda}, {L{\'o}pez
  Fern{\'a}ndez}, {Tadhunter}, {Villar-Mart{\'{\i}}n}, \&
  {Roth}}]{Cortijo2017b}
---. 2017{\natexlab{c}}, \JournalTitle{ArXiv e-prints},
  \href{http://arxiv.org/abs/1706.01896}{{\sffamily arXiv:1706.01896}}

\bibitem[{{Cox} {et~al.}(2006){Cox}, {Dutta}, {Di Matteo}, {Hernquist},
  {Hopkins}, {Robertson}, \& {Springel}}]{Cox2006}
{Cox}, T.~J., {Dutta}, S.~N., {Di Matteo}, T., {et~al.} 2006,
  \href{http://dx.doi.org/10.1086/507474}{\JournalTitle{\apj}, 650, 791}

\bibitem[{{Daddi} {et~al.}(2010){Daddi}, {Elbaz}, {Walter}, {Bournaud},
  {Salmi}, {Carilli}, {Dannerbauer}, {Dickinson}, {Monaco}, \&
  {Riechers}}]{Daddi2010}
{Daddi}, E., {Elbaz}, D., {Walter}, F., {et~al.} 2010,
  \href{http://dx.doi.org/10.1088/2041-8205/714/1/L118}{\JournalTitle{\apjl},
  714, L118}

\bibitem[{{Dekel} \& {Silk}(1986)}]{Dekel1986}
{Dekel}, A., \& {Silk}, J. 1986,
  \href{http://dx.doi.org/10.1086/164050}{\JournalTitle{\apj}, 303, 39}

\bibitem[{{Di Matteo} {et~al.}(2005){Di Matteo}, {Springel}, \&
  {Hernquist}}]{DiMatteo2005}
{Di Matteo}, T., {Springel}, V., \& {Hernquist}, L. 2005,
  \href{http://dx.doi.org/10.1038/nature03335}{\JournalTitle{\nat}, 433, 604}

\bibitem[{{Dopita} {et~al.}(2013){Dopita}, {Sutherland}, {Nicholls}, {Kewley},
  \& {Vogt}}]{Dopita2013}
{Dopita}, M.~A., {Sutherland}, R.~S., {Nicholls}, D.~C., {Kewley}, L.~J., \&
  {Vogt}, F.~P.~A. 2013,
  \href{http://dx.doi.org/10.1088/0067-0049/208/1/10}{\JournalTitle{\apjs},
  208, 10}

\bibitem[{{Driver} {et~al.}(2005){Driver}, {Liske}, {Cross}, {De Propris}, \&
  {Allen}}]{Driver2005}
{Driver}, S.~P., {Liske}, J., {Cross}, N.~J.~G., {De Propris}, R., \& {Allen},
  P.~D. 2005,
  \href{http://dx.doi.org/10.1111/j.1365-2966.2005.08990.x}{\JournalTitle{\mnras},
  360, 81}

\bibitem[{{Duc}(2012)}]{Duc2012}
{Duc}, P.-A. 2012,
  \href{http://dx.doi.org/10.1007/978-3-642-22018-0_37}{\JournalTitle{Astrophysics
  and Space Science Proceedings}, 28, 305}

\bibitem[{{Duc} {et~al.}(2004){Duc}, {Bournaud}, \& {Masset}}]{Duc2004}
{Duc}, P.-A., {Bournaud}, F., \& {Masset}, F. 2004,
  \href{http://dx.doi.org/10.1051/0004-6361:20041410}{\JournalTitle{\aap}, 427,
  803}

\bibitem[{{Efstathiou} {et~al.}(2002){Efstathiou}, {Moody}, {Peacock},
  {Percival}, {Baugh}, {Bland-Hawthorn}, {Bridges}, {Cannon}, {Cole},
  {Colless}, {Collins}, {Couch}, {Dalton}, {de Propris}, {Driver}, {Ellis},
  {Frenk}, {Glazebrook}, {Jackson}, {Lahav}, {Lewis}, {Lumsden}, {Maddox},
  {Norberg}, {Peterson}, {Sutherland}, \& {Taylor}}]{Efstathiou2002}
{Efstathiou}, G., {Moody}, S., {Peacock}, J.~A., {et~al.} 2002,
  \href{http://dx.doi.org/10.1046/j.1365-8711.2002.05215.x}{\JournalTitle{\mnras},
  330, L29}

\bibitem[{{Elmegreen}(2011)}]{Elmegreen2011}
{Elmegreen}, B.~G. 2011, \href{http://dx.doi.org/10.1051/eas/1151004}{in EAS
  Publications Series, Vol.~51, EAS Publications Series, ed. C.~{Charbonnel} \&
  T.~{Montmerle}}, 45

\bibitem[{{Elmegreen} \& {Elmegreen}(2005)}]{Elmegreen2005}
{Elmegreen}, B.~G., \& {Elmegreen}, D.~M. 2005,
  \href{http://dx.doi.org/10.1086/430514}{\JournalTitle{\apj}, 627, 632}

\bibitem[{{Elmegreen} {et~al.}(2009){Elmegreen}, {Elmegreen}, {Fernandez}, \&
  {Lemonias}}]{Elmegreen2009}
{Elmegreen}, B.~G., {Elmegreen}, D.~M., {Fernandez}, M.~X., \& {Lemonias},
  J.~J. 2009,
  \href{http://dx.doi.org/10.1088/0004-637X/692/1/12}{\JournalTitle{\apj}, 692,
  12}

\bibitem[{{Elmegreen} {et~al.}(1993){Elmegreen}, {Kaufman}, \&
  {Thomasson}}]{Elmegreen1993}
{Elmegreen}, B.~G., {Kaufman}, M., \& {Thomasson}, M. 1993,
  \href{http://dx.doi.org/10.1086/172903}{\JournalTitle{\apj}, 412, 90}

\bibitem[{{Fensch} {et~al.}(2016){Fensch}, {Duc}, {Weilbacher}, {Boquien}, \&
  {Zackrisson}}]{Fensch2016}
{Fensch}, J., {Duc}, P.-A., {Weilbacher}, P.~M., {Boquien}, M., \&
  {Zackrisson}, E. 2016,
  \href{http://dx.doi.org/10.1051/0004-6361/201527141}{\JournalTitle{\aap},
  585, A79}

\bibitem[{{Fern{\'a}ndez} {et~al.}(2015){Fern{\'a}ndez}, {Yuan}, {Shen}, {Yin},
  {Chang}, \& {Feng}}]{Fernandez2015}
{Fern{\'a}ndez}, M., {Yuan}, F., {Shen}, S., {et~al.} 2015,
  \href{http://dx.doi.org/10.3390/galaxies3040156}{\JournalTitle{Galaxies}, 3,
  156}

\bibitem[{{Fisher} {et~al.}(2014){Fisher}, {Glazebrook}, {Bolatto},
  {Obreschkow}, {Mentuch Cooper}, {Wisnioski}, {Bassett}, {Abraham},
  {Damjanov}, {Green}, \& {McGregor}}]{Fisher2014}
{Fisher}, D.~B., {Glazebrook}, K., {Bolatto}, A., {et~al.} 2014,
  \href{http://dx.doi.org/10.1088/2041-8205/790/2/L30}{\JournalTitle{\apjl},
  790, L30}

\bibitem[{{Fossati} {et~al.}(2016){Fossati}, {Fumagalli}, {Boselli}, {Gavazzi},
  {Sun}, \& {Wilman}}]{Fossati2016}
{Fossati}, M., {Fumagalli}, M., {Boselli}, A., {et~al.} 2016,
  \href{http://dx.doi.org/10.1093/mnras/stv2400}{\JournalTitle{\mnras}, 455,
  2028}

\bibitem[{{Freedman} {et~al.}(2001){Freedman}, {Madore}, {Gibson}, {Ferrarese},
  {Kelson}, {Sakai}, {Mould}, {Kennicutt}, {Ford}, {Graham}, {Huchra},
  {Hughes}, {Illingworth}, {Macri}, \& {Stetson}}]{Freedman2001}
{Freedman}, W.~L., {Madore}, B.~F., {Gibson}, B.~K., {et~al.} 2001,
  \href{http://dx.doi.org/10.1086/320638}{\JournalTitle{\apj}, 553, 47}

\bibitem[{{Fritz} {et~al.}(2007){Fritz}, {Poggianti}, {Bettoni}, {Cava},
  {Couch}, {D'Onofrio}, {Dressler}, {Fasano}, {Kj{\ae}rgaard}, {Moles}, \&
  {Varela}}]{Fritz2007}
{Fritz}, J., {Poggianti}, B.~M., {Bettoni}, D., {et~al.} 2007,
  \href{http://dx.doi.org/10.1051/0004-6361:20077097}{\JournalTitle{\aap}, 470,
  137}

\bibitem[{{Fritz} {et~al.}(2017){Fritz}, {Moretti}, {Poggianti}, {Gullieuszik},
  {Bruzual}, {Vulcani}, {Nicastro}, {Jaffe'}, {Cervantes Sodi}, {Bettoni},
  {Fasano}, {Charlot}, {Bellhouse}, \& {Hau}}]{Fritz2017}
{Fritz}, J., {Moretti}, A., {Poggianti}, B., {et~al.} 2017, \JournalTitle{ArXiv
  e-prints}, \href{http://arxiv.org/abs/1704.05088}{{\sffamily
  arXiv:1704.05088}}

\bibitem[{{Fu} {et~al.}(2009){Fu}, {Hou}, {Yin}, \& {Chang}}]{Fu2009}
{Fu}, J., {Hou}, J.~L., {Yin}, J., \& {Chang}, R.~X. 2009,
  \href{http://dx.doi.org/10.1088/0004-637X/696/1/668}{\JournalTitle{\apj},
  696, 668}

\bibitem[{{Genzel} {et~al.}(2011){Genzel}, {Newman}, {Jones}, {F{\"o}rster
  Schreiber}, {Shapiro}, {Genel}, {Lilly}, {Renzini}, {Tacconi}, {Bouch{\'e}},
  {Burkert}, {Cresci}, {Buschkamp}, {Carollo}, {Ceverino}, {Davies}, {Dekel},
  {Eisenhauer}, {Hicks}, {Kurk}, {Lutz}, {Mancini}, {Naab}, {Peng},
  {Sternberg}, {Vergani}, \& {Zamorani}}]{Genzel2011}
{Genzel}, R., {Newman}, S., {Jones}, T., {et~al.} 2011,
  \href{http://dx.doi.org/10.1088/0004-637X/733/2/101}{\JournalTitle{\apj},
  733, 101}

\bibitem[{{Hibbard} \& {Vacca}(1997)}]{Hibbard1997}
{Hibbard}, J.~E., \& {Vacca}, W.~D. 1997,
  \href{http://dx.doi.org/10.1086/118603}{\JournalTitle{\aj}, 114, 1741}

\bibitem[{{Hirschmann} {et~al.}(2012){Hirschmann}, {Naab}, {Somerville},
  {Burkert}, \& {Oser}}]{Hirschmann2012}
{Hirschmann}, M., {Naab}, T., {Somerville}, R.~S., {Burkert}, A., \& {Oser}, L.
  2012,
  \href{http://dx.doi.org/10.1111/j.1365-2966.2011.19961.x}{\JournalTitle{\mnras},
  419, 3200}

\bibitem[{{Hopkins} {et~al.}(2013){Hopkins}, {Cox}, {Hernquist}, {Narayanan},
  {Hayward}, \& {Murray}}]{Hopkins2013}
{Hopkins}, P.~F., {Cox}, T.~J., {Hernquist}, L., {et~al.} 2013,
  \href{http://dx.doi.org/10.1093/mnras/stt017}{\JournalTitle{\mnras}, 430,
  1901}

\bibitem[{{Hung} {et~al.}(2014){Hung}, {Sanders}, {Casey}, {Koss}, {Larson},
  {Lee}, {Li}, {Lockhart}, {Shih}, {Barnes}, {Kartaltepe}, \&
  {Smith}}]{Hung2014}
{Hung}, C.-L., {Sanders}, D.~B., {Casey}, C.~M., {et~al.} 2014,
  \href{http://dx.doi.org/10.1088/0004-637X/791/1/63}{\JournalTitle{\apj}, 791,
  63}

\bibitem[{{Izotov} {et~al.}(2006){Izotov}, {Schaerer}, {Blecha}, {Royer},
  {Guseva}, \& {North}}]{Izotov2006}
{Izotov}, Y.~I., {Schaerer}, D., {Blecha}, A., {et~al.} 2006,
  \href{http://dx.doi.org/10.1051/0004-6361:20065622}{\JournalTitle{\aap}, 459,
  71}

\bibitem[{{James} {et~al.}(2010){James}, {Tsamis}, \& {Barlow}}]{James2010}
{James}, B.~L., {Tsamis}, Y.~G., \& {Barlow}, M.~J. 2010,
  \href{http://dx.doi.org/10.1111/j.1365-2966.2009.15706.x}{\JournalTitle{\mnras},
  401, 759}

\bibitem[{{James} {et~al.}(2013{\natexlab{a}}){James}, {Tsamis}, {Barlow},
  {Walsh}, \& {Westmoquette}}]{James2013a}
{James}, B.~L., {Tsamis}, Y.~G., {Barlow}, M.~J., {Walsh}, J.~R., \&
  {Westmoquette}, M.~S. 2013{\natexlab{a}},
  \href{http://dx.doi.org/10.1093/mnras/sts004}{\JournalTitle{\mnras}, 428, 86}

\bibitem[{{James} {et~al.}(2009){James}, {Tsamis}, {Barlow}, {Westmoquette},
  {Walsh}, {Cuisinier}, \& {Exter}}]{James2009}
{James}, B.~L., {Tsamis}, Y.~G., {Barlow}, M.~J., {et~al.} 2009,
  \href{http://dx.doi.org/10.1111/j.1365-2966.2009.15172.x}{\JournalTitle{\mnras},
  398, 2}

\bibitem[{{James} {et~al.}(2013{\natexlab{b}}){James}, {Tsamis}, {Walsh},
  {Barlow}, \& {Westmoquette}}]{James2013b}
{James}, B.~L., {Tsamis}, Y.~G., {Walsh}, J.~R., {Barlow}, M.~J., \&
  {Westmoquette}, M.~S. 2013{\natexlab{b}},
  \href{http://dx.doi.org/10.1093/mnras/stt034}{\JournalTitle{\mnras}, 430,
  2097}

\bibitem[{{Jog}(1997)}]{Jog1997}
{Jog}, C.~J. 1997, \href{http://dx.doi.org/10.1086/304721}{\JournalTitle{\apj},
  488, 642}

\bibitem[{{Kauffmann} {et~al.}(2003){Kauffmann}, {Heckman}, {Tremonti},
  {Brinchmann}, {Charlot}, {White}, {Ridgway}, {Brinkmann}, {Fukugita}, {Hall},
  {Ivezi{\'c}}, {Richards}, \& {Schneider}}]{Kauffmann2003}
{Kauffmann}, G., {Heckman}, T.~M., {Tremonti}, C., {et~al.} 2003,
  \href{http://dx.doi.org/10.1111/j.1365-2966.2003.07154.x}{\JournalTitle{\mnras},
  346, 1055}

\bibitem[{{Kaviraj} {et~al.}(2012){Kaviraj}, {Darg}, {Lintott}, {Schawinski},
  \& {Silk}}]{Kaviraj2012}
{Kaviraj}, S., {Darg}, D., {Lintott}, C., {Schawinski}, K., \& {Silk}, J. 2012,
  \href{http://dx.doi.org/10.1111/j.1365-2966.2011.19673.x}{\JournalTitle{\mnras},
  419, 70}

\bibitem[{{Kennicutt}(1998{\natexlab{a}})}]{Kennicutt1998a}
{Kennicutt}, Jr., R.~C. 1998{\natexlab{a}},
  \href{http://dx.doi.org/10.1146/annurev.astro.36.1.189}{\JournalTitle{\araa},
  36, 189}

\bibitem[{{Kennicutt}(1998{\natexlab{b}})}]{Kennicutt1998b}
---. 1998{\natexlab{b}},
  \href{http://dx.doi.org/10.1086/305588}{\JournalTitle{\apj}, 498, 541}

\bibitem[{{Kewley} \& {Ellison}(2008)}]{Kewley2008}
{Kewley}, L.~J., \& {Ellison}, S.~L. 2008,
  \href{http://dx.doi.org/10.1086/587500}{\JournalTitle{\apj}, 681, 1183}

\bibitem[{{Lagos} {et~al.}(2014){Lagos}, {Papaderos}, {Gomes}, {Smith
  Castelli}, \& {Vega}}]{Lagos2014}
{Lagos}, P., {Papaderos}, P., {Gomes}, J.~M., {Smith Castelli}, A.~V., \&
  {Vega}, L.~R. 2014,
  \href{http://dx.doi.org/10.1051/0004-6361/201323353}{\JournalTitle{\aap},
  569, A110}

\bibitem[{{Lagos} {et~al.}(2009){Lagos}, {Telles}, {Mu{\~n}oz-Tu{\~n}{\'o}n},
  {Carrasco}, {Cuisinier}, \& {Tenorio-Tagle}}]{Lagos2009}
{Lagos}, P., {Telles}, E., {Mu{\~n}oz-Tu{\~n}{\'o}n}, C., {et~al.} 2009,
  \href{http://dx.doi.org/10.1088/0004-6256/137/6/5068}{\JournalTitle{\aj},
  137, 5068}

\bibitem[{{Lagos} {et~al.}(2012){Lagos}, {Telles}, {Nigoche Netro}, \&
  {Carrasco}}]{Lagos2012}
{Lagos}, P., {Telles}, E., {Nigoche Netro}, A., \& {Carrasco}, E.~R. 2012,
  \href{http://dx.doi.org/10.1111/j.1365-2966.2012.21944.x}{\JournalTitle{\mnras},
  427, 740}

\bibitem[{{Larson} \& {Tinsley}(1978)}]{Larson1978}
{Larson}, R.~B., \& {Tinsley}, B.~M. 1978,
  \href{http://dx.doi.org/10.1086/155753}{\JournalTitle{\apj}, 219, 46}

\bibitem[{{Li} {et~al.}(2004){Li}, {Mac Low}, \& {Klessen}}]{Li2004}
{Li}, Y., {Mac Low}, M.-M., \& {Klessen}, R.~S. 2004,
  \href{http://dx.doi.org/10.1086/425320}{\JournalTitle{\apjl}, 614, L29}

\bibitem[{{Liske} {et~al.}(2003){Liske}, {Lemon}, {Driver}, {Cross}, \&
  {Couch}}]{Liske2003}
{Liske}, J., {Lemon}, D.~J., {Driver}, S.~P., {Cross}, N.~J.~G., \& {Couch},
  W.~J. 2003,
  \href{http://dx.doi.org/10.1046/j.1365-8711.2003.06826.x}{\JournalTitle{\mnras},
  344, 307}

\bibitem[{{Magrini} {et~al.}(2007){Magrini}, {V{\'{\i}}lchez}, {Mampaso},
  {Corradi}, \& {Leisy}}]{Magrini2007}
{Magrini}, L., {V{\'{\i}}lchez}, J.~M., {Mampaso}, A., {Corradi}, R.~L.~M., \&
  {Leisy}, P. 2007,
  \href{http://dx.doi.org/10.1051/0004-6361:20077445}{\JournalTitle{\aap}, 470,
  865}

\bibitem[{{Marcon-Uchida} {et~al.}(2010){Marcon-Uchida}, {Matteucci}, \&
  {Costa}}]{Marcon2010}
{Marcon-Uchida}, M.~M., {Matteucci}, F., \& {Costa}, R.~D.~D. 2010,
  \href{http://dx.doi.org/10.1051/0004-6361/200913933}{\JournalTitle{\aap},
  520, A35}

\bibitem[{{Metz} \& {Kroupa}(2007)}]{Metz2007}
{Metz}, M., \& {Kroupa}, P. 2007,
  \href{http://dx.doi.org/10.1111/j.1365-2966.2007.11438.x}{\JournalTitle{\mnras},
  376, 387}

\bibitem[{{Mihos} \& {Hernquist}(1996)}]{Mihos1996}
{Mihos}, J.~C., \& {Hernquist}, L. 1996,
  \href{http://dx.doi.org/10.1086/177353}{\JournalTitle{\apj}, 464, 641}

\bibitem[{{Moreno} {et~al.}(2015){Moreno}, {Torrey}, {Ellison}, {Patton},
  {Bluck}, {Bansal}, \& {Hernquist}}]{Moreno2015}
{Moreno}, J., {Torrey}, P., {Ellison}, S.~L., {et~al.} 2015,
  \href{http://dx.doi.org/10.1093/mnras/stv094}{\JournalTitle{\mnras}, 448,
  1107}

\bibitem[{{Moster} {et~al.}(2011){Moster}, {Macci{\`o}}, {Somerville}, {Naab},
  \& {Cox}}]{Moster2011}
{Moster}, B.~P., {Macci{\`o}}, A.~V., {Somerville}, R.~S., {Naab}, T., \&
  {Cox}, T.~J. 2011,
  \href{http://dx.doi.org/10.1111/j.1365-2966.2011.18984.x}{\JournalTitle{\mnras},
  415, 3750}

\bibitem[{{Okazaki} \& {Taniguchi}(2000)}]{Okazaki2000}
{Okazaki}, T., \& {Taniguchi}, Y. 2000,
  \href{http://dx.doi.org/10.1086/317109}{\JournalTitle{\apj}, 543, 149}

\bibitem[{{Ostriker} \& {Shetty}(2011)}]{Ostriker2011}
{Ostriker}, E.~C., \& {Shetty}, R. 2011,
  \href{http://dx.doi.org/10.1088/0004-637X/731/1/41}{\JournalTitle{\apj}, 731,
  41}

\bibitem[{{Overzier} {et~al.}(2010){Overzier}, {Heckman}, {Schiminovich},
  {Basu-Zych}, {Gon{\c c}alves}, {Martin}, \& {Rich}}]{Overzier2010}
{Overzier}, R.~A., {Heckman}, T.~M., {Schiminovich}, D., {et~al.} 2010,
  \href{http://dx.doi.org/10.1088/0004-637X/710/2/979}{\JournalTitle{\apj},
  710, 979}

\bibitem[{{Poggianti} {et~al.}(2016){Poggianti}, {Fasano}, {Omizzolo},
  {Gullieuszik}, {Bettoni}, {Moretti}, {Paccagnella}, {Jaff{\'e}}, {Vulcani},
  {Fritz}, {Couch}, \& {D'Onofrio}}]{Poggianti2016}
{Poggianti}, B.~M., {Fasano}, G., {Omizzolo}, A., {et~al.} 2016,
  \href{http://dx.doi.org/10.3847/0004-6256/151/3/78}{\JournalTitle{\aj}, 151,
  78}

\bibitem[{{Poggianti} {et~al.}(2017){Poggianti}, {Moretti}, {Gullieuszik},
  {Fritz}, {Jaff{\'e}}, {Bettoni}, {Fasano}, {Bellhouse}, {Hau}, {Vulcani},
  {Biviano}, {Omizzolo}, {Paccagnella}, {D'Onofrio}, {Cava}, {Sheen}, {Couch},
  \& {Owers}}]{Poggianti2017a}
{Poggianti}, B.~M., {Moretti}, A., {Gullieuszik}, M., {et~al.} 2017,
  \href{http://dx.doi.org/10.3847/1538-4357/aa78ed}{\JournalTitle{\apj}, 844,
  48}

\bibitem[{{Prantzos} \& {Boissier}(2000)}]{Prantzos2000}
{Prantzos}, N., \& {Boissier}, S. 2000,
  \href{http://dx.doi.org/10.1046/j.1365-8711.2000.03228.x}{\JournalTitle{\mnras},
  313, 338}

\bibitem[{{Proxauf} {et~al.}(2014){Proxauf}, {{\"O}ttl}, \&
  {Kimeswenger}}]{Proxauf2014}
{Proxauf}, B., {{\"O}ttl}, S., \& {Kimeswenger}, S. 2014,
  \href{http://dx.doi.org/10.1051/0004-6361/201322581}{\JournalTitle{\aap},
  561, A10}

\bibitem[{{Pryke} {et~al.}(2002){Pryke}, {Halverson}, {Leitch}, {Kovac},
  {Carlstrom}, {Holzapfel}, \& {Dragovan}}]{Pryke2002}
{Pryke}, C., {Halverson}, N.~W., {Leitch}, E.~M., {et~al.} 2002,
  \href{http://dx.doi.org/10.1086/338880}{\JournalTitle{\apj}, 568, 46}

\bibitem[{{Recchi} {et~al.}(2007){Recchi}, {Theis}, {Kroupa}, \&
  {Hensler}}]{Recchi2007}
{Recchi}, S., {Theis}, C., {Kroupa}, P., \& {Hensler}, G. 2007,
  \href{http://dx.doi.org/10.1051/0004-6361:20077264}{\JournalTitle{\aap}, 470,
  L5}

\bibitem[{{Renaud} {et~al.}(2015){Renaud}, {Bournaud}, \& {Duc}}]{Renaud2015}
{Renaud}, F., {Bournaud}, F., \& {Duc}, P.-A. 2015,
  \href{http://dx.doi.org/10.1093/mnras/stu2208}{\JournalTitle{\mnras}, 446,
  2038}

\bibitem[{{Renaud} {et~al.}(2014){Renaud}, {Bournaud}, {Kraljic}, \&
  {Duc}}]{Renaud2014}
{Renaud}, F., {Bournaud}, F., {Kraljic}, K., \& {Duc}, P.-A. 2014,
  \href{http://dx.doi.org/10.1093/mnrasl/slu050}{\JournalTitle{\mnras}, 442,
  L33}

\bibitem[{{Rupke} {et~al.}(2010){Rupke}, {Kewley}, \& {Barnes}}]{Rupke2010}
{Rupke}, D.~S.~N., {Kewley}, L.~J., \& {Barnes}, J.~E. 2010,
  \href{http://dx.doi.org/10.1088/2041-8205/710/2/L156}{\JournalTitle{\apjl},
  710, L156}

\bibitem[{{S{\'a}nchez} {et~al.}(2012){S{\'a}nchez}, {Kennicutt}, {Gil de Paz},
  {van de Ven}, {V{\'{\i}}lchez}, {Wisotzki}, {Walcher}, {Mast}, {Aguerri},
  {Albiol-P{\'e}rez}, {Alonso-Herrero}, {Alves}, {Bakos}, {Bart{\'a}kov{\'a}},
  {Bland-Hawthorn}, {Boselli}, {Bomans}, {Castillo-Morales}, {Cortijo-Ferrero},
  {de Lorenzo-C{\'a}ceres}, {Del Olmo}, {Dettmar}, {D{\'{\i}}az}, {Ellis},
  {Falc{\'o}n-Barroso}, {Flores}, {Gallazzi}, {Garc{\'{\i}}a-Lorenzo},
  {Gonz{\'a}lez Delgado}, {Gruel}, {Haines}, {Hao}, {Husemann},
  {Igl{\'e}sias-P{\'a}ramo}, {Jahnke}, {Johnson}, {Jungwiert}, {Kalinova},
  {Kehrig}, {Kupko}, {L{\'o}pez-S{\'a}nchez}, {Lyubenova}, {Marino},
  {M{\'a}rmol-Queralt{\'o}}, {M{\'a}rquez}, {Masegosa}, {Meidt},
  {Mendez-Abreu}, {Monreal-Ibero}, {Montijo}, {Mour{\~a}o}, {Palacios-Navarro},
  {Papaderos}, {Pasquali}, {Peletier}, {P{\'e}rez}, {P{\'e}rez}, {Quirrenbach},
  {Rela{\~n}o}, {Rosales-Ortega}, {Roth}, {Ruiz-Lara},
  {S{\'a}nchez-Bl{\'a}zquez}, {Sengupta}, {Singh}, {Stanishev}, {Trager},
  {Vazdekis}, {Viironen}, {Wild}, {Zibetti}, \& {Ziegler}}]{Sanchez2012}
{S{\'a}nchez}, S.~F., {Kennicutt}, R.~C., {Gil de Paz}, A., {et~al.} 2012,
  \href{http://dx.doi.org/10.1051/0004-6361/201117353}{\JournalTitle{\aap},
  538, A8}

\bibitem[{{Schawinski} {et~al.}(2009){Schawinski}, {Virani}, {Simmons}, {Urry},
  {Treister}, {Kaviraj}, \& {Kushkuley}}]{Schawinski2009}
{Schawinski}, K., {Virani}, S., {Simmons}, B., {et~al.} 2009,
  \href{http://dx.doi.org/10.1088/0004-637X/692/1/L19}{\JournalTitle{\apjl},
  692, L19}

\bibitem[{{Schlafly} \& {Finkbeiner}(2011)}]{Schlafly2011}
{Schlafly}, E.~F., \& {Finkbeiner}, D.~P. 2011,
  \href{http://dx.doi.org/10.1088/0004-637X/737/2/103}{\JournalTitle{\apj},
  737, 103}

\bibitem[{{Schoenmakers} {et~al.}(1997){Schoenmakers}, {Franx}, \& {de
  Zeeuw}}]{Schoenmakers1997}
{Schoenmakers}, R.~H.~M., {Franx}, M., \& {de Zeeuw}, P.~T. 1997,
  \href{http://dx.doi.org/10.1093/mnras/292.2.349}{\JournalTitle{\mnras}, 292,
  349}

\bibitem[{{Searle} \& {Zinn}(1978)}]{Searle1978}
{Searle}, L., \& {Zinn}, R. 1978,
  \href{http://dx.doi.org/10.1086/156499}{\JournalTitle{\apj}, 225, 357}

\bibitem[{{Sengupta} {et~al.}(2014){Sengupta}, {Scott}, {Dwarakanath},
  {Saikia}, \& {Sohn}}]{Sengupta2014}
{Sengupta}, C., {Scott}, T.~C., {Dwarakanath}, K.~S., {Saikia}, D.~J., \&
  {Sohn}, B.~W. 2014,
  \href{http://dx.doi.org/10.1093/mnras/stu1463}{\JournalTitle{\mnras}, 444,
  558}

\bibitem[{{Spergel} {et~al.}(2007){Spergel}, {Bean}, {Dor{\'e}}, {Nolta},
  {Bennett}, {Dunkley}, {Hinshaw}, {Jarosik}, {Komatsu}, {Page}, {Peiris},
  {Verde}, {Halpern}, {Hill}, {Kogut}, {Limon}, {Meyer}, {Odegard}, {Tucker},
  {Weiland}, {Wollack}, \& {Wright}}]{Spergel2007}
{Spergel}, D.~N., {Bean}, R., {Dor{\'e}}, O., {et~al.} 2007,
  \href{http://dx.doi.org/10.1086/513700}{\JournalTitle{\apjs}, 170, 377}

\bibitem[{{Tempel} {et~al.}(2012){Tempel}, {Tago}, \&
  {Liivam{\"a}gi}}]{Tempel2012}
{Tempel}, E., {Tago}, E., \& {Liivam{\"a}gi}, L.~J. 2012,
  \href{http://dx.doi.org/10.1051/0004-6361/201118687}{\JournalTitle{\aap},
  540, A106}

\bibitem[{{Teyssier} {et~al.}(2010){Teyssier}, {Chapon}, \&
  {Bournaud}}]{Teyssier2010}
{Teyssier}, R., {Chapon}, D., \& {Bournaud}, F. 2010,
  \href{http://dx.doi.org/10.1088/2041-8205/720/2/L149}{\JournalTitle{\apjl},
  720, L149}

\bibitem[{{Toomre}(1977)}]{Toomre1977}
{Toomre}, A. 1977, in Evolution of Galaxies and Stellar Populations, ed. B.~M.
  {Tinsley} \& R.~B.~G. {Larson}, D.~Campbell, 401

\bibitem[{{Torrey} {et~al.}(2012){Torrey}, {Cox}, {Kewley}, \&
  {Hernquist}}]{Torrey2012b}
{Torrey}, P., {Cox}, T.~J., {Kewley}, L., \& {Hernquist}, L. 2012,
  \href{http://dx.doi.org/10.1088/0004-637X/746/1/108}{\JournalTitle{\apj},
  746, 108}

\bibitem[{{Vazdekis} {et~al.}(2010){Vazdekis}, {S{\'a}nchez-Bl{\'a}zquez},
  {Falc{\'o}n-Barroso}, {Cenarro}, {Beasley}, {Cardiel}, {Gorgas}, \&
  {Peletier}}]{Vazdekis2010}
{Vazdekis}, A., {S{\'a}nchez-Bl{\'a}zquez}, P., {Falc{\'o}n-Barroso}, J.,
  {et~al.} 2010,
  \href{http://dx.doi.org/10.1111/j.1365-2966.2010.16407.x}{\JournalTitle{\mnras},
  404, 1639}

\bibitem[{{V{\'e}ron-Cetty} \& {V{\'e}ron}(2010)}]{Veron2010}
{V{\'e}ron-Cetty}, M.-P., \& {V{\'e}ron}, P. 2010,
  \href{http://dx.doi.org/10.1051/0004-6361/201014188}{\JournalTitle{\aap},
  518, A10}

\bibitem[{{Weinberg}(1995)}]{Weinberg1995}
{Weinberg}, M.~D. 1995,
  \href{http://dx.doi.org/10.1086/309803}{\JournalTitle{\apjl}, 455, L31}

\bibitem[{{Wetzstein} {et~al.}(2007){Wetzstein}, {Naab}, \&
  {Burkert}}]{Wetzstein2007}
{Wetzstein}, M., {Naab}, T., \& {Burkert}, A. 2007,
  \href{http://dx.doi.org/10.1111/j.1365-2966.2006.11360.x}{\JournalTitle{\mnras},
  375, 805}

\bibitem[{{White} \& {Rees}(1978)}]{White1978}
{White}, S.~D.~M., \& {Rees}, M.~J. 1978,
  \href{http://dx.doi.org/10.1093/mnras/183.3.341}{\JournalTitle{\mnras}, 183,
  341}

\bibitem[{{Wild} {et~al.}(2014){Wild}, {Rosales-Ortega}, {Falc{\'o}n-Barroso},
  {Garc{\'{\i}}a-Benito}, {Gallazzi}, {Gonz{\'a}lez Delgado}, {Bekerait{\'e}},
  {Pasquali}, {Johansson}, {Garc{\'{\i}}a Lorenzo}, {van de Ven}, {Pawlik},
  {Per{\'e}z}, {Monreal-Ibero}, {Lyubenova}, {Cid Fernandes},
  {M{\'e}ndez-Abreu}, {Barrera-Ballesteros}, {Kehrig}, {Iglesias-P{\'a}ramo},
  {Bomans}, {M{\'a}rquez}, {Johnson}, {Kennicutt}, {Husemann}, {Mast},
  {S{\'a}nchez}, {Walcher}, {Alves}, {Aguerri}, {Alonso Herrero},
  {Bland-Hawthorn}, {Catal{\'a}n-Torrecilla}, {Florido}, {Gomes}, {Jahnke},
  {L{\'o}pez-S{\'a}nchez}, {de Lorenzo-C{\'a}ceres}, {Marino},
  {M{\'a}rmol-Queralt{\'o}}, {Olden}, {del Olmo}, {Papaderos}, {Quirrenbach},
  {V{\'{\i}}lchez}, \& {Ziegler}}]{Wild2014}
{Wild}, V., {Rosales-Ortega}, F., {Falc{\'o}n-Barroso}, J., {et~al.} 2014,
  \href{http://dx.doi.org/10.1051/0004-6361/201321624}{\JournalTitle{\aap},
  567, A132}

\bibitem[{{Zackrisson} \& {Riehm}(2010)}]{Zackrisson2010}
{Zackrisson}, E., \& {Riehm}, T. 2010,
  \href{http://dx.doi.org/10.1155/2010/735284}{\JournalTitle{Advances in
  Astronomy}, 2010, 735284}

\end{thebibliography}


\end{document}